\documentclass[aps,prb,twocolumn,showpacs]{revtex4}
\usepackage{amsmath,amssymb,epsf,bm}
\usepackage{graphicx}
\usepackage{srcltx}
\begin{document}

\title{Aharonov-Casher oscillations of spin current through
a multichannel mesoscopic ring}
\author{V.V. Shlyapin, A.G. Mal'shukov}
\affiliation{Institute of Spectroscopy, Russian Academy of
Sciences,
142190 Troitsk, Moscow region, Russia}
\begin{abstract}
The Aharonov-Casher (AC) oscillations of spin current through a
2D ballistic ring in the presence of Rashba spin-orbit
interaction and external magnetic field has been calculated
using
the semiclassical path integral method. For classically chaotic
trajectories the Fokker-Planck equation determining dynamics of
the particle spin polarization has been derived. On the basis 
of this equation an analytic expression for the spin conductance 
has been
obtained taking into account a finite width of the ring arms
carrying large number of conducting channels. It was shown that
the finite width results in a broadening and damping of spin 
current AC oscillations. We found that an external magnetic 
field leads to appearance of new nondiagonal components of 
the spin conductance,
allowing thus by applying a rather weak magnetic field to change a
direction of the transmitted spin current polarization.
\end{abstract}
 \maketitle

\section{\label{Introduction}Introduction}

The Aharonov-Casher effect is a spectacular demonstration of
the fundamental role of the spin-orbit interaction (SOI) in
electronic transport. This effect is a non-Abelian analog of
the Aharonov-Bohm effect. An electronic wave entering a two
dimensional ring is splitted into two waves traveling trough 
the upper and the
lower arms and interfering on the exit from the ring. 
Due to SOI the
spinor components in each of the waves obtain the relative phase 
shift. This shift, in its turn, gives rise to a destructive or
constructive interference pattern in the transmittance 
probability,
that results in a number of oscillation effects on electron
transport parameters. Notably, that in semiconductor
heterostructures SOI can be varied through the gate voltage
manipulation, suggesting interesting opportunities for
practical applications of this effect in spintronics.

A simplest model to study the AC effect is a one-dimensional
(1D) or one-channel ring with none or few scatterers 
of electrons. Aronov
and Lyanda-Geller \cite{Aronov_LG} studied oscillations of
the magnetoconductance in a 1D ring in the presence of Rashba
\cite{Bychkov} SOI. Ya-Sha Yi \textit{et. al.} \cite{Ya_Sha}
considered a joint effect of the Zeeman coupling, magnetic flux
and SOI on the conductance of a 1D ring beyond the adiabatic
approximation employed in Ref.~\onlinecite{Aronov_LG}. 
Nitta \textit{et.
al.} \cite{Nitta_theory} noted that SOI alone, without the
magnetic
field, can cause oscillations of the ring electric conductance.
\mbox{Meijer} et al.\cite{Meijer_1D_Hamiltonian} pointed out
that the
Hamiltonian used earlier in Ref.~\onlinecite{Aronov_LG} was not
quite
correct and, in particular, was not Hermitian. Using the
correct Hamiltonian Frustaglia and Richter\cite{Frustaglia}
revised the
expression for the conductance found in
Ref.~\onlinecite{Nitta_theory}.

In disordered mesoscopic systems the AC effect is strongly
modified.
Multiple impurity scatterings lead to averaging out of some
strong
oscillations that were presented in ideal 1D rings. There is a
similarity to AB effect where the fundamental $h/e$ peak in the
Fourier spectrum of magnetoconductance vanishes and is
substituted
for $h/2e$ Al'tshuler, Aronov, and Spivak \cite{AAS}
oscillations.
Mathur and Stone \cite{Mathur} have demonstrated that a similar
weak
localization effect takes place in the case of AC oscillations
in a
thin diffusive ring with Rashba SOI. They have shown that the
average conductance periodically varies as a function of the
SOI coupling constant with a period that is 1/2 of the ideal 
ring
conductance oscillations. This prediction have been confirmed
experimentally by Koga et al.\cite{Koga_et_al} who observed 
that the
amplitude of $h/2e$ magnetoconductance oscillations (AAS type
oscillations) oscillates itself with varying the applied gate
voltage and, correspondingly, the SOI strength. On the other
hand,
the fundamental AB, as well as AC oscillations show up in
mesoscopic conductance  fluctuations. The
conductance correlator has been calculated in
Ref.~\onlinecite{Malsh_Diffus}
for a diffusive ring in the presence of
Zeeman coupling and Rashba SOI. It was found that the amplitude
of
$h/e$ AB peak shows oscillations with varying SOI, similar to
those
that have been observed for the $h/2e$ peak in
Ref.~\onlinecite{Koga_et_al}.

Although 1D, as well as diffusive models are useful to
elucidate fundamental physical effects associated with SOI,
they can not be fully applied to realistic semiconductor systems
used in experiments. For example, an attempt to interpret the
observed
\cite{Morpurgo} splitting of the AB power spectrum within the
diffusion model \cite{Malsh_Diffus,Engel_Loss} was not
successful.
In Ref. \onlinecite{Morpurgo}, as well as in other experimental
works \cite{Koga_et_al,Yau_Shayegan}, the mesoscopic loops
carry many channels. At the same time their sizes are comparable
to the electron mean free path. Therefore, neither of the above
theoretical models can be applied. In this situation an
approach based on path integrals along classic trajectories can
be
fruitful. Such a method has been applied to transport in
mesoscopic systems in a number of works
\cite{Gutz,JBS_chaos,Berry_Robnik,JBS_Blum_Smil}. It
was also employed for calculation of the spin conductance
through
a classically chaotic and regular cavities and rings with
Rashba SOI \cite{Malsh,Cheng_Hung}. In Ref.\onlinecite{Malsh} it
was done analytically by applying the method of trajectory
averaging, while numerical simulations of path integrals 
were performed in Ref.\onlinecite{Cheng_Hung}. 
In both cases pronounced AC
oscillations of the spin conductance with varying SOI constant
had
been found. Their important distinction from oscillations of
the electric conductance is that they appear in the main
semiclassical
approximation, not involving weak localization or other quantum
corrections.

While in a multichannel loop studied in Ref. \onlinecite{Malsh}
the electron motion was two-dimensional, the AC phase
accumulation
had an effectively one-dimensional character.
The reason is that in a
thin enough loop SOI causes only a small variation of spinor
amplitudes during particle motion along any straight segment of a
trajectory. In the leading approximation ignoring AC phase
fluctuations associated with a finiteness of a loop width, a
phase evolution depends only on a coordinate along the loop and
finite width effects vanish. On the other hand, as follows from
the Monte Carlo analysis \cite{Cheng_Hung}, when a particle
lifetime within the ring is long enough, the finite width effect 
on the amplitude and shape of AC oscillations is strong.

In order to elucidate this problem we will calculate the spin
conductance taking into
account the finite width effect in a classically chaotic
multi-channel ring with the Rashba spin-orbit interaction and a
uniform magnetic field applied perpendicular to the ring. A
chaotic motion of particles can be provided by non ideal
boundaries
of the ring, as well as by random potential variations inside
it.
Starting from an analog of the Landauer formula derived for the
spin conductance in Appendix~\ref{Landauer}, we apply the path
integral method to this conductance. Within this approximation
spin
dependent transmission amplitudes for each of the classical
paths
decompose into a product of a spin independent transmission
amplitude and a matrix determining evolution of spinor
components
along this trajectory. The expression for the spin conductance
that
is quadratic in these amplitudes should be double summed over
the
trajectories afterwards. Ignoring weak localization and other
quantum corrections, only the terms diagonal with respect to
the trajectories are then retained. As a result, the spin
conductance
takes a form of a $3 \times 3$ matrix averaged over the
trajectories. Based on known statistical properties of chaotic
trajectories we will derive Fokker-Planck equations describing
spin
evolution of a particle moving through the ring and
analytically calculate the average of spin conductance matrix
components. The finite
width effects will be analyzed that show up in an additional
broadening and decreasing of AC oscillations. The magnetic
field, in
its turn, results in appearance of spin conductance components
that
were equal to zero in the absence of the field, suggesting thus
an opportunity to rotate the spin polarization on the exit from 
the ring.

The outline of the paper is as follows. Section II contains a
description of the model system we used in our theory. In
section
III an expression for the spin conductance is obtained in the
form
of a sum over classical trajectories. In section IV the
Fokker-Planck equation for the spin polarization distribution
function is derived. On the basis of this equation the spin
conductance averaged over chaotic trajectories is calculated.
The
discussion of results is presented in Section V.

\section{\label{Statement}The model}
\begin{figure}[bp]
\includegraphics[width=247pt]{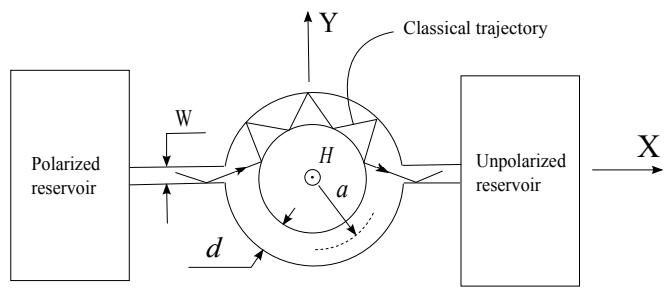}
\caption{ Geometry of the problem under consideration. The
$Z$-axis points to the reader. 
$W$~is the leads width,
$d~$is the ring arms width, and $a$ is the radius 
of the ring.
} \label{fig1}
\end{figure}

We consider spin transport through a 2D ring which is connected
via
two symmetrically placed leads to two reservoirs of electrons
(see
Fig.~\ref{fig1}) and is subject to the magnetic field
perpendicular
to the ring plane. The latter gives rise to the Zeeman
interaction
\begin{equation}
\label{eq_H_Z} \delta H_Z=\frac{\hbar \omega_H}{2}\, \sigma_z\, ,
\hphantom{w}
\omega_H=\frac{g}{2}\frac{|e|H}{m^*c}\, ,
\end{equation}
where $e$ is the electron charge, $H$ is the magnetic field, 
$g$ is the g-factor, $m^*$ is the effective electron mass,
and $c$ is the light velocity. 

The Rashba spin-orbit interaction is assumed to take place only
in
the range of the ring. It has the form
\begin{equation}
\label{eq_H_R} \delta H_R=\alpha_{so}\,
\hat{\bm{\sigma}}\cdot[\hat{{\mathbf p}}\times{\mathbf z}]\, .
\end{equation}
Here $\alpha_{so}$ is the SOI constant, the vector
$\hat{\bm{\sigma}}$ consists of the Pauli matrices
$\hat{\sigma}_x$,
$\hat{\sigma}_y$ and $\hat{\sigma}_z$, $\hat{\mathbf p}$ is the
momentum operator and ${\mathbf z }$ is a unit vector parallel
to the $Z$-axis. 
We assume the hard wall reflection of electrons from
the ring boundaries. Since the particle spin is conserved upon
such a reflection, Eqs. (\ref{eq_H_Z}) and (\ref{eq_H_R})
determine a spin dynamics in the ring. 

The reservoirs are assumed to be in a  local thermodynamic
equilibrium with a given polarization (magnetization). For
simplicity we assume that the left reservoir is polarized
(along a unit vector $\bm \nu$), while
the right one is unpolarized. The polarization of the left
reservoir is characterized by the chemical potentials
$\mu_{L\uparrow}=\mu_L+\frac{\delta\mu}{2}$ and
$\mu_{L\downarrow}=\mu_L-\frac{\delta\mu}{2}$, for spin-up and
spin-down
(relative to $\bm \nu$)
electrons, respectively. For the right reservoir we assume, in
its
turn, that $\mu_{R\uparrow}=\mu_{R\downarrow}=\mu_{L}$. This
situation is of particular interest for our further analysis,
because establishing of thermodynamic equilibrium between
spin subsystems in
the reservoirs will be accompanied by the spin current, while
the
electric current will be absent. In the linear response regime
the
spin current in the right lead can be expressed (see Appendix~A) 
as
\begin{equation}\label{J}
J_j=\sum_i g_{ji}\nu_i\delta \mu\, ,
\end{equation}
where $i$ and $j$ take the values $x,y,z$, and $\nu_j$ is
the $j$-th component of $\bm \nu$.

\section{\label{Semiclas}Semiclassical approximation}
We assume that the leads connecting the ring with reservoirs
are ideal conductors with a constant cross section. This suggests
the use of the Landauer approach for calculation of the spin
current. 
The latter, however, has an important distinction 
from the electric
current, because it does not conserve in the region with SO
interaction. 
Let us assume, for example, that such a region 
is ideally transparent. Even in this case the transmitted 
spin current will not be equal to the incident one. 
Alternatively, we may consider a polarized reservoir 
connected via {\it one} lead to a region with SO interaction. 
In the stationary case the electric current through this lead 
will be zero. At the same time , the spin current will be 
finite, because spin polarizations of incident and reflected 
electrons can be different. This sort of spin transport has 
been  considered in Ref.~\onlinecite{Malsh} where a ring 
played the role of the
region with SO interaction. 
   
So, we see that 
Landauer formula should be modified to take into account 
effects of nonconcerving spin. 
In general case it is done 
in Appendix~\ref{Landauer}. For a particular
set up considered in this work the expression for the spin
conductance takes a simple form, somewhat similar to the
Landauer formula
\begin{equation}
\label{sp_cond_def_copy}
g_{ji}=\frac{1}{4\pi \hbar}\sum_{pp'}
{\rm Tr}\left\{
\hat{t}_{p'p}^{\, +}(\mu_R)\,\hat{\sigma_j}\,
\hat{t}_{p'p}(\mu_R)\,\hat{\sigma_i}
\right\}
\end{equation}

Here
$\hat{t}_{p'p}(\mu_R)$ is the $2\times 2$ matrix composed of the
transmission amplitudes
$t_{p'p}^{\uparrow\uparrow}(\mu_R)$,$\,t_{p'p}^{\uparrow\downarrow}(\mu_R)\,$,
$t_{p'p}^{\downarrow\uparrow}(\mu_R)\,$,
$t_{p'p}^{\downarrow\downarrow}(\mu_R)\,$ from the channel $p$
of
the left lead to the channel $p'$ of the right lead. The
summation
is performed over open channels. Note, that this formula
implies that for calculation of all spin conductance components
it is not sufficient to calculate the spin-resolved 
transmission coefficients
$T^{\alpha' \alpha}\equiv\sum_{p'p}|t_{p'p}^{\alpha'
\alpha}|^2$ (studied, for example, in
Refs.~\onlinecite{Frustaglia, Frustaglia_2}). 
This is readily seen, for example, by inspecting the
expression
\begin{equation*}
g_{xz}=\frac{1}{4 \pi \hbar}\sum_{pp'}
\left\{
(t_{p'p}^{\downarrow \uparrow})^*\,t_{p'p}^{\uparrow \uparrow}
-(t_{p'p}^{\uparrow \downarrow} )^*\,t_{p'p}^{\downarrow
\downarrow}
\right\},
\end{equation*}
which follows from (\ref{sp_cond_def_copy}).

To find the transmission amplitudes in (\ref{sp_cond_def_copy})
we
will follow Ref.~\onlinecite{FL} (see also
Ref.~\onlinecite{Kaw})
for the spinless transmission amplitude at the Fermi energy
$E_F$.
In the presence of spin degrees of freedom it allows the
evident generalization:
\begin{equation}
\label{eq_t_G} t^{\alpha'\alpha}_{p'p}=-i\hbar\sqrt{v_{p'}
v_p}\int dy
dy' u_{p'}^*(y')u_p(y)G(y'\alpha',y\alpha|E_F) \, .
\end{equation}
$v_{p'}$ ($v_p$) is the longitudinal 
velocity in the
$p'$\nobreakdash-th ($p$\nobreakdash-th) channel of the right
(left) lead. $G$ is the retarded Green's function. The
integration
is performed over cross-sections of the leads. We assume that
the
leads and ring arms are wide enough, so that there are many
channels below the Fermi energy. This allows to apply the
semiclassical approximation (see e.g. Ref.
\onlinecite{JBS_chaos})
to Eq. (\ref{eq_t_G}). Within this approximation the path
integral
expression for the Green's function is replaced by the sum of
amplitudes corresponding to classical trajectories traversing
the
ring (details of this procedure may be found in
Ref.~\onlinecite{Gutz}). Integration over $y$ and $y'$ is
performed by the stationary phase method (large parameter in
the exponent is the number of open channels in the leads). The
result
is
\begin{equation}
\label{eq_t_semic}
t^{\alpha'\alpha}_{p'p}=
\sum_{s}t_0(s)S_s^{\alpha'\alpha} \, .
\end{equation}
The label $s$ enumerates the trajectories that 
enter the ring at the angle
$\theta=\pm\sin^{-1} \left(p\pi/k_F W\right)$ relative to the
$x-$axis and
exit it at $\theta'=\pm\sin^{-1}\left(p'\pi/k_F W\right)$, 
where $W$ and $k_F$ are the leads width and 
Fermi wave number, respectively.
$t_0(s)$ is the spin independent transmission amplitude
corresponding to the $s-$th classical trajectory,
\begin{multline}
\label{t_0_s}
t_0(s)=-\sqrt{\frac{i}{2N_m}}
\;\mathrm{sgn}\left[\theta_s\right]
\;\mathrm{sgn}\left[\theta'_s\right]\sqrt{\tilde{A}_s}\\
\times
\exp\left\{\vphantom{\oint}i k_F L_s + i k_F \left(\sin \theta_s
y_{s}-\sin\theta_s'y'_{s}\right)\right.\\
\left.+\frac{ie}{\hbar c}\oint_s
\bm{A}d\bm{r}-i\tilde{\mu}_s\frac{\pi}{2}\right\}
\, .
\end{multline}
Here $N_m\equiv int\{k_F W/\pi\}$ is the number of open
channels,
$L_s$ is the length of the $s$-th trajectory, $y_s$ ($y_s'$)
stands
for the $y$-coordinate of the entrance (exit) point of $s$-th
trajectory, $\bm{A}$ is the vector potential corresponding to
the
magnetic field applied to the ring. Other quantities in
(\ref{t_0_s}) are
\begin{gather*}
\tilde{A}_s=\frac{1}{a\cos \theta_s'}\left|\frac{\partial
y(\theta,\theta')}{\partial\theta'}\right|\, ,\\
\tilde{\mu}_s=\mu_s + \Theta\left[\frac{\partial
\theta(y,y')}{\partial y}\right] + \Theta\left[-\frac{\partial
\theta'(\theta,y')}{\partial y'}\right]\, ,
\end{gather*}
where $\mu_s$ is the Maslov index \cite{JBS_chaos,Gutz} and
$\Theta$ is the Heaviside step function. 
The matrix $S_s^{\alpha'\alpha}$ in
(\ref{eq_t_semic}) determines an evolution of the spin state
along $s$-th trajectory. It should be noted that Eq.
(\ref{eq_t_semic})
has been derived assuming that classical trajectories do not
depend
on the spin dynamics. This allowed to write the terms entering
into
the sum in Eq. (\ref{eq_t_semic}) in the form of a product of
spin
dependent and spin independent parts. In fact, this assumption
means
that in the leading semiclassical approximation we ignore a
difference between Fermi velocities corresponding to spin-split
subbands. It can be done, if during the time between two
consecutive
collisions with ring boundaries, a divergence of two wave
packets
belonging to these subbands will be much less than the electron
wavelength. The corresponding condition can be written as
$L_{SO}=\hbar/\alpha_{SO}\ m^* \gg d$ where $L_{SO}$ is the
spin-orbit length that measures the SOI strength and $d$ is the
ring
width.

For a spin dependent Hamiltonian consisting of the two terms
represented by Eqs. (\ref{eq_H_R}) and (\ref{eq_H_Z}), the
evolution operator can be expressed as
\begin{equation}
\label{expr_S} \widehat{S}_s={\cal T}
\left[\exp\left\{-\frac{i}{\hbar} \int dt \,
\left(\alpha_{so}p_F \,\hat{\bm{\sigma}}\cdot\bm{n}_t +
\frac{\hbar\,\omega_H}{2}\,\hat{\sigma}_z
\right)\right\}\right]\, .
\end{equation}
Here ${\cal T}$ is the time ordering symbol, 
$p_F$ is the Fermi momentum,
$\bm{n}_t$ is the unit vector parallel to $\mathbf{p}\times
\mathbf{z}$, and $\textbf{p}$ ($|\textbf{p}|=p_F$) is the
electron momentum.
Note, that direction $\bm{n}_t$ of the effective magnetic field
generated by the SO interaction changes
its sign when a particle reverses its motion direction.

Substituting (\ref{eq_t_semic}) into (\ref{sp_cond_def_copy})
we obtain
\begin{equation}
\label{eq_g_semic} g_{ji}=\frac{1}{4 \pi \hbar} \sum_{p',p}
\sum_{s, u}
t_0^*(s)t_0(u)
{\rm Tr}\left\{\hat{\sigma}_i \widehat{S}_s^{+}\hat{\sigma}_{j}
\widehat{S}_u\right\} \, .
\end{equation}
In this equation each semiclassical amplitude $t_0(s)$ contains
a
phase factor $\exp\left\{2\pi i L_s/\lambda\right\}$. Since the
path lengths $L_s$ and $L_u$ of trajectories in
Eq.~(\ref{eq_g_semic}) are much longer than the electron
wavelength $\lambda$, the terms with $s\neq u$ oscillate
rapidly even with a small variation of the particle energy, as
well as
slight change of the loop shape and/or impurity positions. In
an experimental situation it can also be gate voltage variations
and
magnetic field switching \cite{Morpurgo}. On the other hand,
the diagonal terms with $s=u$ do not oscillate. If one is not
interested in mesoscopic fluctuations of the spin conductance,
or
quantum corrections to it, only the terms with $s=u$ have to be
retained \cite{JBS_Blum_Smil}. On the basis of the ergodic
hypothesis of Ref.~\onlinecite{Lee_Stone_Fukuyama} this
procedure
may also be treated as averaging over a random ensemble of the
rings.

Thus, from (\ref{eq_g_semic}) the averaged spin conductance
$\langle
g_{ij}\rangle$ is obtained as
\begin{equation}\label{gij_av}
\langle g_{ij}\rangle = \frac{1}{2 \pi \hbar}\sum_s|t_0(s)|^2 K_{ij}^s\ .
\end{equation}
where
\begin{equation}\label{Kij}
K_{ij}^s =\frac{1}{2} {\rm Tr}\left\{\hat{\sigma}_i
\widehat{S}_s\hat{\sigma}_j \widehat{S}_s^{\dag}
\right\}\ .
\end{equation}
The electrical conductance $G$ after such averaging takes the
form
\begin{equation}
\label{G} \langle G \rangle=(e/\pi \hbar)\sum_s|t_0(s)|^2\, ,
\end{equation}
where the factor $2$ accounts for the spin degrees of freedom.
It
should be noted that $\langle G \rangle$ given by (\ref{G}) does
not
depend on the SO interaction strength. On the other hand,
oscillations of $G$ with the varying Rashba constant have
been predicted in Ref.~\onlinecite{Aronov_LG}. This distinction
can be explained by importance of quantum effects in an ideal
1D ring considered in Ref.~\onlinecite{Aronov_LG}, while such
effects are small in a disordered multichannel system that 
we study here.
They can be taken into account as weak localization corrections
to AC oscillations similar to those studied for diffusive
rings. \cite{Mathur}

Let us now consider the transparency
\[
N=\sum_{p',p}\left|\,t_{p'p}\,\right|^2
\]
for a spinless particle. Here $t_{p'p}$  is the
transmission amplitude from the channel $p$ on the left to the
channel $p'$ on the right. So, the transparency $N$ is
normalized
to the number of open channels $N_m$. Applying configurational
averaging to the above expression we obtain it as a sum over
trajectories:
\begin{equation}
\label{N_cl}
\langle N \rangle=N_{cl}\equiv \sum_s \left|t_0(s)\right|^2\, .
\end{equation}
Hence, the ratio
\[
P(s)=\frac{\left|t_0(s)\right|^2}{N_{cl}}
\]
can be identified with the probability that an electron chooses
the $s$-th trajectory to pass the ring. Expressing
$\left|t_0(s)\right|^2$ from the last formula and substituting
it
into (\ref{gij_av}) we obtain
\begin{equation}\label{gij_av_1}
\langle g_{ij}\rangle= g_0\langle K_{ij}^s\rangle_s \, ,
\end{equation}
where $\langle \cdot \rangle_s$ denotes averaging over trajectories with the
probability $P(s)$ and 
$2e^2 g_0=e^2 N_{cl}/ \pi \hbar$
is a classical  conductance of the ring \cite{Kaw} .

We assume that the classical motion of an electron inside the
ring
is {\it chaotic}. The chaotic dynamics is provided by small
scale
bumps and other irregularities on the ring boundaries, while
macroscopically the ring preserves its regular shape. For
chaotic
trajectories after long enough time an electron "forgets"
through
which of the leads it entered the ring. Together with an
assumption that the
leads are symmetric this allows to conclude that probability for a
particle to be reflected is equal to the probability to be
transmitted. So, $N_{cl}=N_m/2$ and $g_0=N_m/4\pi\hbar$.

Formula (\ref{gij_av_1}) will be used for calculation of the
spin
conductance $\langle g_{ij}\rangle$. However, another physically
more
transparent representation of $\langle g_{ij}\rangle$ can be
suggested. As shown in Appendix~\ref{transformation}, $K_{ij}^s$is
the $i$-th component of the electron polarization at the end of
the
$s$-th trajectory, provided that at its beginning the electron
was
polarized along the $j$-th axis. Taking this into account one
can
introduce the effective polarization vector
$\bm{P}_{eff}(s|\bm{e}_j)\equiv
\left[g(s)/g_0\right]\left\{K_{xj}^s,K_{yj}^s,K_{zj}^s \right\}$ of
electrons at the end of the $s$-th trajectory. Here
$g(s)=\left|t_0(s)\right|^2/2\pi\hbar$ is the flux of spinless
particles that pass the ring through the $s$-th trajectory. Eq.
(\ref{gij_av})
then takes the form
\begin{equation}
\label{gij_av_phys}
\langle g_{ij}\rangle = g_0 \sum_s P^i_{eff}(s|\bm{e}_j)\, .
\end{equation}
From comparison of (\ref{gij_av_1}) and (\ref{gij_av_phys}) we
see
that $\langle K_{ij}^s\rangle_s$ may be treated as $i$-th
component
of the effective polarization vector $\bm{P}_{eff}(\bm{e}_j)$
on the exit from the ring,
\begin{equation}
\label{P_eff_full}
\langle K_{ij}^s\rangle_s=P^i_{eff}(\bm{e}_j)\equiv\sum_s
P^i_{eff}(s|\bm{e}_j)\, .
\end{equation}
Equation (\ref{gij_av_phys}) means that 
the spin current on the
exit from the ring is given by the sum of effective
polarizations
corresponding to each classical trajectory, times the flux of
particles passing the ring. From this fact an immediate
conclusion
follows: if the resonance condition is satisfied, that is the
effective polarizations corresponding to different trajectories
on
the exit from the ring are co-directional, then the spin current
will have a maximum.

\section{\label{Calc}Fokker-Planck equation}

According to Eq.~(\ref{gij_av_1}) calculation of the spin 
conductance $\langle g_{ij} \rangle$ reduces to averaging of the polarization
transformation matrix $K_{ij}^s$ over trajectories. To perform
this averaging the distribution function 
of $K_{ij}^s$ is needed. One of the standard ways 
to find it is the  following. 
We note that the quantities $K_{ij}^s$ in (\ref{Kij})
correspond to the end of the $s$-th trajectory. It is
evident however that one may extend definition (\ref{Kij})
to any time instant $t$ on the 
trajectory. Taking then a time derivative of (\ref{Kij})
we arrive at the stochastic differential equation
(of the Langevin type) for $K_{ij}^s$. 
This equation may be used to calculate
drift and diffusion coefficients in the 
Fokker-Planck equation. Solving then this Fokker-Planck
equation one can find the desired distribution function.

\subsection{\label{Din_eq_S_matrix}Dynamical equation for
the spin S-matrix along a trajectory}

As a preparation step to our calculation 
we note that at $H=0$ 
eq. (\ref{expr_S}) 
for $\widehat{S}_s$ can be written 
in the  
form of a contour integral over the $s$-th trajectory:
\begin{equation}
\label{expr_S_H0} \widehat{S}_s(H=0)= 
{\cal T}\left[\exp\left\{-\frac{i}{L_{SO}}
\oint_s d{\bm l} \,
\cdot\left[\bm{z}\times\hat{\bm{\sigma}}\right]
\right\}\right]\, ,
\end{equation}
As any unitary operator acting in the spin space, $S_s(H=0)$ can
also 
be written in the form of the rotation operator
\begin{equation}
\label{Phase}
\widehat{S}_s(H=0)=\exp\left\{i\frac{\hat{\sigma}_{\bm{N}}\Psi}{2}\right\} \, ,
\end{equation}
where $\hat{\sigma}_{\bm{N}}$ is a projection of $\hat{\bm{\sigma}}$
onto
some unit vector $\bm{N}$. An important consequence of
(\ref{expr_S_H0}) is that $\bm{N}$ and $\Psi$ depend only on
the geometry of the $s$-th trajectory and don't depend on the
dynamics
of motion along the trajectory. This is the reason why
$\Psi$ may be called a geometric phase\cite{Malsh}.

\begin{figure}
\includegraphics[width=247pt]{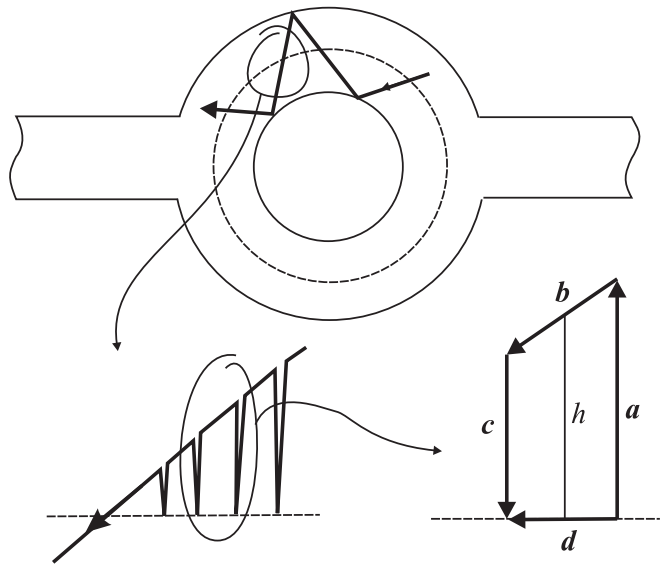}
\caption{Replacement of the real trajectory with the fictitious
one in the case of the ring of finite width. Both trajectories
give rise to the same evolution operator $S_s$ in the absence of the
magnetic field.} \label{fig2}
\end{figure}

To extend the analysis of Ref.~\onlinecite{Malsh} 
and take into account finiteness of the
ring width we 
divide the time interval $(0, t)$ into small
subintervals $\Delta t_i=t_i-t_{i-1};\, t_0=0<t_1<\dots<t_n=t$.
After that $\widehat{S}$ can be represented in the form 
(the index $s$ is omitted when it doesn't lead to 
any confusion)
\begin{equation}
\label{S_product}
\widehat{S}(t)=\widehat{S}(\Delta t_n)
\widehat{S}(\Delta t_{n-1})\dots 
\widehat{S}(\Delta t_1)\, .
\end{equation}
The real trajectory is transformed next by adding to each i-th
segment a path passed in direct and opposite directions, as
shown in
Fig.~\ref{fig2}. It is seen from (\ref{expr_S_H0}), that paths
passed twice in the opposite directions don't contribute to
$\widehat{S}_s(H=0)$. Consequently, each term 
$\widehat{S}(\Delta t_i)$ in
(\ref{S_product}) may be replaced without changing 
$\widehat{S}_s$ with
\begin{equation}
\label{S_abc}
\widehat{S}_{\bm{abc}}\equiv \widehat{S}_{\bm{c}}(H=0)\widehat{S}_{\bm{b}}\widehat{S}_{\bm{a}}
(H=0)\, .
\end{equation}
Here $\bm b$ is the trajectory segment corresponding to the
time interval $\Delta t_i$, $\bm a$ and $\bm c$ connect the
middle
line
of the ring with $\bm b$, as explained in Fig.~\ref{fig2}.
Expanding exponents in (\ref{expr_S}) and (\ref{expr_S_H0}) one
obtains
\begin{equation}
\label{S_abc_expan}
\widehat{S}_{\bm{abc}}=1-\frac{i}{L_{SO}}\hat{\sigma}_r 
dl
-i\frac{\omega_H}{2}\hat{\sigma}_z \Delta t_i
+2\frac{i}{L_{SO}^2}\hat{\sigma}_z 
\Sigma_{\bm{abc,-d}}\, ,
\end{equation}
where $dl=|\bm{d}|$ for counterclockwise 
rotation of the electron
and $dl=-|\bm{d}|$ otherwise. $\hat{\sigma}_r$ 
is the projection of
$\hat{\bm{\sigma}}$ on the radius vector. 
$\Sigma_{\bm{abc,-d}}$
is the oriented area of the trapezium 
formed by the vectors 
$\bm{a, b, c}$ and $\bm{-d}$. 
It is positive if the contour $\bm{abc,-d}$ is positively 
oriented and negative otherwise. 
The last term in (\ref{S_abc_expan}) takes into
account the finite width of the ring. 
It leads to a phase proportional to the
area embraced by the trajectory, analogously to the finite
width effect on the Aharonov-Bom phase\cite{Kaw_Nak}.

Formulae (\ref{S_product}) and (\ref{S_abc_expan})
yield the following dynamical equation for the spin
evolution operator
\begin{equation}
\label{eq_S}
\frac{\partial \widehat{S}}{\partial t}=\left\{i\gamma \dot{\phi}
\left(e^{i\phi}\hat{\sigma}_{-}+e^{-i\phi}\hat{\sigma}_{+}\right)
-i\eta\hat{\sigma}_z\right\}\widehat{S} \, .
\end{equation}
Here
\begin{gather}
\gamma=a/L_{SO}\, ,\label{gamma}\\
\hat{\sigma}_{\pm}=\frac{\hat{\sigma}_x \pm i\hat{\sigma}_y}{2}
\, ,\nonumber\\
\eta \equiv \eta(t) =
\frac{\omega_H}{2}-2\gamma^2\frac{h(t)}{a}\dot{\phi}\, ,
\label{eta}
\end{gather}
$a$ is the ring radius (distance from the center to the middle
line
of the ring), and $\dot{\phi}$ is the time derivative of the
polar
angle $\phi$ counted from the negative direction of the
$OX$-axis.
The deviation $h(t)$ of the electron trajectory from the middle
line
(see Fig.~\ref{fig2}) is taken positive for points outside the
circle formed by this line and negative otherwise.

Dynamical equation (\ref{eq_S}) simplifies if we 
change the system of coordinates. 
First we perform transformation to the system
$\widetilde{OXYZ}$ rotating together with the electron
($\widetilde{OZ}$-axis coincides with the $OZ$-axis,
$\widetilde{OX}$ and $\widetilde{OY}$ axes rotate with the
angular
velocity $\dot{\phi}$ around $OZ$-axis). In this system the
evolution operator takes the form
\begin{gather}
\label{S_rotated} \widetilde{S}=\widehat{R}_z(t) \widehat{S}
\widehat{R}^{-1}_z(0) \, ,\\
\widehat{R}_z(t)=\exp\left\{i\hat{\sigma}_z\phi(t)/2\right\}\,
.\label{R_z}
\end{gather}
Using (\ref{eq_S}) and (\ref{S_rotated}) and taking into
account that $\widehat{R}^{-1}_z(0)=1$ one can verify that $\widetilde{S}$
satisfies
the equation
\begin{equation}
\label{motion_S_rotated}
\frac{d\,\widetilde{S}}{d\,t}=\left\{
-i\frac{\hat{\sigma}_{\theta}\omega_0}{2}-i\eta\hat{\sigma}_z
\right\}\widetilde{S}\, ,
\end{equation}
where $\hat{\sigma}_{\theta}$ is a projection of $\hat{\bm{\sigma}}$
onto the vector
\begin{gather}
\bm{\theta}=\left\{2\gamma/\zeta,0,1/\zeta\right\}\,
,\label{vector_theta}\\
\zeta=\sqrt{1+4\gamma^2} \, , \label{zeta}
\end{gather}
and
\begin{equation}
\label{omega_0}
\omega_0=-\zeta\dot{\phi}\, .
\end{equation}

Next, we rotate $\widetilde{OXYZ}$ around $\widetilde{OY}$-axis,until
$\widetilde{OZ}$ comes parallel to $\bm{\theta}$. The new
coordinate system denoted as $\widetilde{OX'YZ'}$ will be
called below the tilted rotating (TR) system. In TR coordinates
the
evolution operator takes the form
\begin{gather}
\widetilde{S'}=\widehat{R}_y 
\widetilde{S} \widehat{R}_y^{-1}=\widehat{R}_y \widehat{R}_z 
\widehat{S} \widehat{R}_y^{-1}
\label{S_tr}\\
\widehat{R}_y=\exp\left\{i\hat{\sigma}_y\chi_0/2\right\}\, .\nonumber
\end{gather}
The angle $\chi_0$ between $\widetilde{OZ}$ and
$\widetilde{OZ'}$ is
defined by the relations
\begin{equation}
\label{chi_0}
\cos\chi_0=1/\zeta, \, \sin\chi_0=2\gamma/\zeta\, .
\end{equation}
The evolution of $\widetilde{S'}$ is determined by
\begin{gather}
\frac{d\,\widetilde{S'}}{d\,t}=i\frac{\hat{\bm{\sigma}}\bm{\Omega}}{2}
\widetilde{S'} \label{dS_dt_TR}\\
\bm{\Omega}\equiv\left\{\frac{4\gamma}{\zeta}\eta,\, 0,\,
-\omega_0-\frac{2\eta}{\zeta}\right\} \, .\label{Omega}
\end{gather}
The physical meaning of the TR system is rather transparent. At
vanishing ring width and zero magnetic field this is the
coordinate
system where the effective magnetic field produced by the Rashba SOI
is parallel to the $\widetilde{OZ'}$-axis.

\subsection{\label{Dynamics_of_spin_polarization} 
Evolution of spin state along a trajectory in terms
of polarization vector}
Transition from $\widehat{S}$ in (\ref{Kij}) to $\widetilde{S'}$ yields
\begin{equation}
\label{K_ij_TR}
K_{ij}=\frac{1}{2}{\rm Tr}\left\{\widetilde{\sigma}'_i 
\,\widetilde{[\sigma_j(t)]}'\right\}\,
,\end{equation}
where
\begin{gather}
\widetilde{\sigma}'_i 
= \widehat{R}_y 
\widehat{R}_z \hat{\sigma}_i 
\widehat{R}_z^+ 
\widehat{R}_y^+\, ,\label{sigma_i_TR}\\
\widetilde{[\sigma_j(t)]}'
=\widetilde{S'} \hat{\sigma}'_j \widetilde{S'}^+ 
=\widehat{R}_y \widehat{R}_z 
\hat{\sigma}_j(t) 
\widehat{R}_z^+ 
\widehat{R}_y^+
\, ,\label{sigma_j_t_TR}\\
\hat{\sigma}_j'=\widehat{R}_y \hat{\sigma}_j
\widehat{R}_y^{-1}=
\left.\tilde{\sigma}_j'\right|_{t=0} \, ,\nonumber \\
\hat{\sigma}_j(t)
=\widehat{S} 
\hat{\sigma}_j \widehat{S}^+ \nonumber\, .
\end{gather}
It is easy to see that $\widetilde{\sigma}'_i$ in
(\ref{sigma_i_TR})
is an image of $\hat{\sigma}_i$ after 
transformation to the TR coordinates. 
Analogously, $\widetilde{[\sigma_j(t)]}'$ 
is is an image of $\hat{\sigma}_j(t)$ after the same 
transformation. Note, that both $\widetilde{\sigma}'_i$
and $\widetilde{[\sigma_j(t)]}'$
depend on time: $\widetilde{\sigma}'_i$ --- due to rotation of
the TR system of coordinates, $\widetilde{[\sigma_j(t)]}'$  
--- due to rotation of the electron spin (described by the evolution
operator) and rotation of the TR system of coordinates.
Further, both $\widetilde{\sigma}'_i$ and 
$\widetilde{[\sigma_j(t)]}'$
can be decomposed into the sums over $\hat{\sigma}_p$,
\begin{gather}
\widetilde{\sigma}'_ i= \sum_{p=1}^3 \Lambda^0_{pi}\hat{\sigma}_p \,
\label{sigma_i_TR_dec}\\
\widetilde{[\sigma_j(t)]}'=
\sum_{p=1}^3 \Lambda_{pj}\hat{\sigma}_p \label{sigma_j_t_TR_dec}\, .
\end{gather}
The superscript $"0"$ at $\Lambda_{pi}$ in (\ref{sigma_i_TR_dec})
indicates, that $\Lambda_{pi}$ transforms into $\Lambda^0_{pi}$
with $\alpha_{so}\longrightarrow 0$.
The unit matrix $\hat{\sigma}_0$ is not present in these sums
because the traces of $\widetilde{\sigma}'_i$ and 
$\widetilde{[\sigma_j(t)]}'$ are zero, as
can be seen from (\ref{sigma_i_TR}) and (\ref{sigma_j_t_TR}).
Using (\ref{eq_dens_matr}) 
from Appendix~\ref{transformation} one can check, that
$\Lambda_{pj}$ are TR coordinates of the electron spin, provided that
the initial spin was $\bm{e}_j$. Analogously, $\Lambda^0_{pi}$  are
coordinates of $\bm{e}_i$ in the TR system. After substitution
into (\ref{K_ij_TR}) the sums
(\ref{sigma_i_TR_dec}) and (\ref{sigma_j_t_TR_dec})
lead to $K_{ij}$ written in the form of scalar
product,
\begin{gather}
K_{ij}=\bm{\Lambda}^0_i\cdot\bm{\Lambda}_j\, ,\label{K_ij_scalar}\\
\bm{\Lambda}^0_i\equiv\left\{\Lambda^0_{1i},
\Lambda^0_{2i},\Lambda^0_{3i}\right\}\, ,
\nonumber\\
\bm{\Lambda}_j\equiv\left\{ \Lambda_{1j},\Lambda_{2j},
\Lambda_{3j}\right\}\, .\nonumber
\end{gather}
The components of $\bm{\Lambda}^0_i$ in (\ref{K_ij_scalar}) being the
coordinates of $\bm{e}_i$ in the TR system are defined only by
its
orientation with respect to the original system, that is by the
angles $\phi$ and $\chi_0$. The angle $\phi$ at the end of the
trajectory is $\phi=\pi+2\pi n$, where $n=\pm 1, \pm2 ...$ is
the
winding number. The angle $\chi_0$ is also fixed, see
(\ref{chi_0}).
Hence, components of $\bm{\Lambda}^0_i$ 
at the end of a trajectory
are constants defined explicitly by
Eq. (\ref{sigma_i_TR}):
\begin{subequations}
\label{Lambda_0_i}
\begin{align}
&\bm{\Lambda}^0_{x}=\left\{-\frac{1}{\zeta},0,-\frac{2\gamma}{\zeta}\right\}\,
,\\
&\bm{\Lambda}^0_{y}=\left\{0,-1,0\right\}\, ,\\
&\bm{\Lambda}^0_{z}=\left\{-\frac{2\gamma}{\zeta},0,\frac{1}{\zeta}\right\}\,
.
\end{align}
\end{subequations}
With constant $\bm{\Lambda}^0_i$ averaging of $K_{ij}$ over
trajectories reduces to averaging of $\bm{\Lambda}_j$. 
We shall perform this averaging
with the use of the distribution function $\cal P$ of
$\bm{\Lambda}_j$.
This function will be obtained from the Fokker-Planck equation
which will be derived and solved below.

The equation of motion for $\widetilde{[\sigma_j(t)]}'$ 
is found from (\ref{sigma_j_t_TR}) and (\ref{dS_dt_TR}). 
In view of 
the expansion (\ref{sigma_j_t_TR_dec}) 
it can be written as an equation of motion 
for $\bm{\Lambda}_j$:
\begin{equation}
\label{Lambda_j_eq}
\frac{d\bm{\Lambda}_j}{d\,
t}=\left[\bm{\Lambda}_j\times\bm{\Omega}\right]\, .
\end{equation}
This equation should be supplemented with the initial
conditions
\begin{subequations}
\label{Lambda_init_cond}
\begin{align}
&\left.\bm{\Lambda}_x\right|_{t=o}=\left\{1/\zeta,\, 0,\,
2\gamma/\zeta\right\}\, , \\
&\left.\bm{\Lambda}_y\right|_{t=o}=\left\{0,\, 1,\,
0\right\}\, , \\
&\left.\bm{\Lambda}_z\right|_{t=o}=\left\{-2\gamma/\zeta,\, 0,\,
1/\zeta\right\}\, ,
\end{align}
\end{subequations}
which can be derived from (\ref{sigma_j_t_TR}) and (\ref{sigma_j_t_TR_dec}).

\subsection{\label{Stochastic_diff_eqs}Stochastic differential
equations for the angles determining the position of an 
electron and direction of its spin polarization}
It follows from (\ref{sigma_j_t_TR_dec}) that $\bm{\Lambda}_j^2=
{\rm Tr}\left\{ \widetilde{[\sigma_j(t)]}'^{\, 2} \right\}/2=1$. 
This allows to describe $\bm{\Lambda}_j$
by only two variables, the polar angle $\Phi$ and 
azimuthal angle $\Theta$. 
Eq.(\ref{Lambda_j_eq}) is then reduced to
\begin{equation*}
\left\{
\begin{array}{rl}
\dot{\Theta}&=\frac{4\gamma}{\zeta}\eta\sin\Phi\\
\dot{\Phi}&=\omega_0(t)+\frac{2}{\zeta}\eta\left[1+2\gamma\cos\Phi\cot\Theta\right]
\, .
\end{array}
\right.
\end{equation*}
It is convenient to represent the angle $\Phi$ in the form
$\Phi=\psi+\delta$, where
\begin{equation}
\label{psi} \psi\equiv-\zeta\phi \, .
\end{equation}
For these new variables we obtain the system of equations:
\begin{equation}
\label{stoch_syst}
\left\{
\begin{array}{rl}
\dot{\Theta}&=2\gamma\left(\frac{\omega_H}{\zeta}+\omega_W(t)\right)\sin(\psi+\delta)\\
\dot{\delta}&=\left(\frac{\omega_H}{\zeta}+\omega_W(t)\right)
\left[1+2\gamma\cos(\psi+\delta)\cot\Theta\right]\\
\dot{\psi}&=\omega_0(t)
\, ,
\end{array}
\right.
\end{equation}
where we have introduced the frequency $\omega_W$ associated
with
the finite width of the ring,
\begin{equation}
\label{omega_W}
\omega_W(t)=\mu(t)\omega_0(t), \,
\mu(t)=\left(\frac{2\gamma}{\zeta}\right)^2\frac{h(t)}{a}\, .
\end{equation}
At the weak enough magnetic field $\omega_H\ll\omega_0$. If, in
addition, the ring is narrow, $h(t)\ll a$, then $\mu(t)\ll 1$
and
$\omega_W\ll\omega_0$, as follows from (\ref{omega_W}).
Returning to
(\ref{stoch_syst}) we thus see that $\Theta$ and $\delta$ are
"slow"
variables, while $\psi$ is a "fast" variable. Such a separation
of
variables simplifies considerably the Fokker-Planck equation,
which
will be derived below from the system of stochastic
differential equations (\ref{stoch_syst}).

\subsection{\label{Fokker_Planck_equation} Derivation 
of the Fokker-Planck equation}
Parameters of the Fokker-Planck equation for the distribution
function ${\cal P}(\Theta,\delta,\psi)$ are determined by the
drift
$A_{\Theta}$, $A_{\delta}$, $A_{\psi}$ and diffusion $B_{ij};\,
i,j=\left\{\Theta, \delta, \psi\right\}$ coefficients
\cite{Drift_Diffus}. Let us first consider  the diffusion
coefficient $B_{\psi \psi}$ given by
\begin{equation}
\label{B_psi_psi1} B_{\psi \psi}= \lim_{\Delta t \rightarrow 0}
\frac{\overline{\left(\Delta \psi\right)^2}}{\Delta t} \,,
\end{equation}
where $\Delta \psi$ is the increment of the stochastic process
$\psi(t)$. It follows from Eq. (\ref{psi}) that
$\overline{(\Delta
\psi)^2}= \zeta^2\overline{(\Delta \phi)^2}$. On the other hand,it
was shown in Ref.~{\onlinecite{Berry_Robnik}} that the winding
number $w$ of the classically chaotic trajectories has a
Gaussian
distribution
\begin{equation}
\label{P_W_T}
{\cal P}\left(w\left|T\right.\right)=
\left(2\pi T/T_1\right)^{-1/2}\exp\left\{\frac{-w^2
}{2T/T_1}\right\}.
\end{equation}
where the constant $T_1$ is the characteristic time of one
turn, $\langle w^2(T_1)\rangle =1$. This equation means that the
winding
of
trajectories is a diffusion process. One can extend
(\ref{P_W_T}) to
a range of $w=\phi/2\pi <1$ assuming that a particle advances
diffusively along a ring arm. Such situation takes place if
the rotation direction $\dot{\phi}/|\dot{\phi}|$ changes many
times,
while passing the angular distance $\Delta \phi < 2\pi$.
Moreover,
the angular distance between two consecutive changes of a
rotation
direction must be small enough to ensure a small change of
$\Phi$
and $\psi$. Hence, as follows from Eq. (\ref{psi}) this
distance 
must be $\ll\zeta^{-1}$. We assume that scattering from
ring
boundaries and spatially fluctuating potential make this
condition
satisfied.

From (\ref{B_psi_psi1}) and (\ref{P_W_T}) with $w=\phi/2\pi$ we
immediately find
\begin{equation}
\label{B_psi_psi} B_{\psi \psi}\equiv \lim_{\Delta t \rightarrow0}
\frac{\overline{\left(\Delta \psi\right)^2}}{\Delta t}
=\frac{\left(2\pi\zeta\right)^2}{T_1} \,,
\end{equation}
where the overline denotes averaging over trajectories. Other
diffusion coefficients, as well as drift coefficients, are
conveniently expressed in terms of the diffusion coefficient of
the
auxiliary stochastic processes $u(t)$, which is defined by its
stochastic differential
\begin{equation}
du=\omega_W(t)dt\label{du}\, .
\end{equation}
An assumption that the deviation $h(t)$ of a trajectory from
the middle line of the ring and the angle $\phi$ fluctuate
independently
leads to the absence of correlations between the stochastic
processes $\psi(t)$ and $u(t)$. Assuming also a uniform
distribution
of $h$ over the width $d$ of the ring arms, we find from Eq.
(\ref{omega_W})
\begin{gather}
\label{B_psi_u} B_{u \psi}=B_{\psi u}\equiv \lim_{\Delta t
\rightarrow 0}
\frac{\overline{\Delta \psi\,\Delta u}}{\Delta t}=0\, , \\
\label{B_uu} B_{uu}\equiv \lim_{\Delta t \rightarrow 0}
\frac{\overline{\left(\Delta u\right)^2}}{\Delta t}=
\left(\frac{2\gamma}{\zeta}\right)^4\frac{d^{\,2}}{12a^2}B_{\psi\psi}\,
.
\end{gather}

Further, Eq. (\ref{stoch_syst}) can be used to express the
small increments $\Delta \Theta, \Delta \delta, \Delta \psi$ in
the form of integrals
over time interval
$\Delta t$. Then, after averaging procedure the
limits $\Delta t \rightarrow 0$ must be taken. For example,
$A_{\Theta}=
  \lim_{\Delta t \rightarrow 0}
  \overline{\Delta \Theta}/ \Delta t, \,
B_{\Theta\delta}= \lim_{\Delta t \rightarrow 0}
\overline{\Delta\Theta\,\Delta \delta}/ \Delta t$. We thus
obtain
\begin{equation}
\label{A}
\left\{
\begin{array}{l}
A_{\Theta}=\omega_H\frac{2\gamma}{\zeta}\sin\left(\psi+\delta\right)
+\frac{B_{uu}}{2}2\gamma\cos\left(\psi+\delta\right) F\, ,\\
A_{\psi}=0\, ,\\ A_{\delta}=\frac{\omega_H}{\zeta}F-
\frac{B_{uu}}{2}2\gamma\sin\left(\psi+\delta\right)
\left[cot\Theta\right.\\
\left.\phantom{aaaaaaaaaaaaaaaa}+2\gamma\frac{\cos^2\Theta+1}{\sin^2\Theta}
\cos\left(\psi+\delta\right)\right]\, ,
\end{array}
\right.
\end{equation}
\begin{equation}
\label{B}
\left\{
\begin{array}{l}
B_{\Theta\Theta}=\sin^2(\psi+\delta)(2\gamma)^2
B_{uu}\, ,\\
B_{\delta\delta}=F^2
B_{uu}\, ,\\
B_{\Theta\delta}=2\gamma\sin\left(\psi+\delta\right)
F B_{uu}
\, ,\\
B_{\psi\Theta}=B_{\psi\delta}=0\, ,
\end{array}
\right.
\end{equation}
where
\begin{align*}
&F=1+2\gamma\cos(\psi+\delta)\cot\Theta \, .
\end{align*}
These coefficients should be inserted in the Fokker-Planck
equation
\begin{multline*}
\frac{\partial}{\partial t} \left({\cal P} \sin
\Theta\right)=\\-\sum_{i}\frac{\partial}{\partial x_i}
\left(A_{x_i}{\cal P}
\sin
\Theta\right)
+\frac{1}{2}\sum_{i,\,j}\frac{\partial^2}{\partial x_i\partial x_j} 
\left(B_{x_i\,x_j}{\cal P} \sin \Theta\right)\,
,
\end{multline*}
where the variables $x_i\:(i=1,2,3)$ denote $\Theta,\,
\delta,\,\psi$. The probability density $\cal P$ is normalized
in such a
way
that an integral of ${\cal P}\sin\Theta$ over $\Theta,\delta$
and
$\psi$ is 1. In this way we arrive at the following equation
\begin{multline}
\label{FP_cumbersome} 
\sin\Theta 
\frac{\partial {\cal P}}{\partial t}
= \frac{B_{uu}}{2}(2\gamma)^2\frac{1}{\sin \Theta}
\frac{1+\cos\left[2(\psi+\delta)\right]}{2}{\cal P}\\
-\omega_H\frac{2\gamma}{\zeta}\sin(\psi + \delta)
\left(\frac{\partial {\cal P}}{\partial \Theta}\sin \Theta +
{\cal P}\cos\Theta\right) +\dots\\+
\frac{B_{\psi\psi}}{2}\sin \Theta 
\frac{\partial^{\,2}{\cal P}}{\partial \psi^{\,2}}\, ,
\end{multline}
where the dots stand for other terms that are proportional to
$B_{uu}$
or $\omega_H$. All the terms in r.h.s, except for the last one,
do not
contain derivatives over ${\psi}$. For a narrow enough ring and
weak
magnetic field,  $B_{uu}$ and $\omega_H\ll B_{\psi\psi}$. So,
diffusion in the space of the "slow" variables $\Theta$ and
$\delta$
is indeed slower than diffusion through the "fast" variable
$\psi$.
Hence, the diffusion equation can be averaged over the fast
variable. After averaging (\ref{FP_cumbersome}) over $\psi$ we
arrive at
the Fokker-Planck equation of the form
\begin{multline}
\label{FP_averaged} \frac{\partial {\cal P}}{\partial t} =
-\frac{\omega_H}{\zeta}\frac{\partial {\cal P}}{\partial
\delta}+
B_{uu}\gamma^2\frac{1}{\sin \Theta}\frac{\partial}{\partial
\Theta} \left(\sin\Theta\frac{\partial {\cal
P}}{\partial\Theta}\right)\\
 +B_{uu}\left(\frac{1}{2}+\gamma^2
cot^2\Theta\right) \frac{\partial^2{\cal P}}{\partial\delta^2}
+\frac{B_{\psi\psi}}{2}\frac{\partial^2{\cal
P}}{\partial\psi^2}\,
.
\end{multline}
This equation should be solved together with the initial
conditions
for each of the vectors $\bm{\Lambda}_j$.
\begin{equation}
\label{P_init_cond} {\cal P}(\Theta, \psi, \delta|t=0)=
\delta(\cos\Theta-\cos\Theta_0^j)
\delta(\delta-\delta_0^j)\delta(\psi)\, ,
\end{equation}
where the angles $\Theta_0^j$, $\delta_0^j$ defining the
initial positions of the vectors $\bm{\Lambda}_j$ are found from
(\ref{Lambda_init_cond}). Using Eq.(\ref{chi_0}) $\theta_0$ and
$\delta_0$ can be expressed in terms of $\chi_0$. The initial
value
of $\psi$ should be zero since it is proportional to the
initial value of $\phi$, which is zero.

\subsection{\label{Solution_Focker_Plank}
Solution of the Focker-Plank equation}
After Laplace transformation with respect to time and Fourier
transformation with respect to $\delta$ and $\psi$ the equation
(\ref{FP_averaged}) is reduced to the ordinary differential
equation
\begin{multline}
\label{FP_Laplace_Fourier} \lambda
v-\left\{\frac{1}{\sin\Theta}\frac{\partial}{\partial \Theta}
\left(\sin\Theta\frac{\partial}{\partial\Theta}\right)
-\frac{\kappa^2}{\sin^2\Theta}\right\}v\\
=\delta\left(\cos\Theta-\cos\Theta_0^j\right)\, ,
\end{multline}
where
\begin{gather*}
v=\gamma^2 B_{uu}e^{i \kappa \delta_0^j}\tilde{\cal P}\, ,\\
\tilde{\cal P}\equiv\tilde{\cal P}(\Theta,\kappa,q|p)
=\int d\psi d\delta\, e^{-i(\kappa\delta+q\psi)}\\
\times\int_0^{+\infty}dt\, e^{-p\,t} {\cal
P}(\Theta,\delta,\psi|t)\,
\\ \lambda=\frac{p+i\kappa\frac{\omega_H}{\zeta} +\kappa^2
B_{uu}\left(\frac{1}{2}-\gamma^2\right)
+q^2\frac{B_{\psi\psi}}{2}}{\gamma^2 B_{uu}}\, .
\end{gather*}
The solution of Eq. (\ref{FP_Laplace_Fourier}) can be expressed
in
terms of eigenfunctions of the linear operator in l.h.s. of
this equation. In our case they are the associated
Legendre functions $P_n^{|\kappa|}(\cos\Theta)$ and we obtain
\begin{multline}
\label{P} {\cal P}(\Theta,\delta,\psi|t)
=\frac{e^{-\psi^2/(2B_{\psi\psi}t)}}{\sqrt{2\pi
B_{\psi\psi}t}}\sum_{\kappa=-\infty}^{+\infty}
\frac{e^{ik(\delta-\delta_0^j-\omega_H t/\zeta)}}
{2\pi}\\
\times\sum_{n=|\kappa|}^{+\infty}\frac{2n+1}{2}\frac{(n-|\kappa|)!}{(n+|\kappa|)!}
P_n^{|\kappa|}(\cos\Theta)P_n^{|\kappa|}(\cos\Theta_0^j)\\
\times
e^{-B_{uu}\left[|\kappa|^2\left(\frac{1}{2}-\gamma^2\right)+\gamma^2n(n+1)\right]t}\,
,
\end{multline}

In this equation only a factor in front of the first sum
depends on $\psi$. It is clear that this factor determines a
probability distribution of $\psi$. At the end of trajectories
(at the exit from the ring), when
$\phi=2\pi w$, $w=\pm 1/2, \pm 3/2 ...$,
it coincides with the winding number distribution (\ref{P_W_T}). 
The remaining part of (\ref{P}) is evidently the conditional 
(for given $\psi$) probability distribution of
$\delta$ and $\Theta$. This function can be used for averaging 
of the polarization vectors $\bm{\Lambda}_j$ over trajectories 
with a given winding number $w$. 

\subsection{\label{Averaging}
Averaging of the polarization vectors $\bm{\Lambda_j}$}
Since the unit vectors $\bm{\Lambda}_j$ in Eq.
(\ref{Lambda_j_eq}) are defined by their respective polar and
azimuthal
angles $\Theta$ and $\Phi$, one can calculate easily their
average values using Eq. (\ref{P}) and taking into account that
$\Phi=\psi+\delta$. We thus arrive at the following expressions
for
the averages $\langle \bm{\Lambda}_j\rangle |_{\psi,T}$ at fixed
trajectory
duration $T$ and a given $\psi$ (winding number),
\begin{subequations}
\label{Lambda_av}
\begin{align}
&\langle \Lambda_{xj}\rangle|_{\psi,T}=\sin\Theta_0^j\cos\left(\psi+\frac{\omega_H}{\zeta}T+\delta_0^j\right)
e^{-T/ \tau_{\perp}}\, ,\\
&\langle \Lambda_{yj}\rangle|_{\psi,T}=\sin\Theta_0^j\sin\left(\psi+\frac{\omega_H}{\zeta}T+\delta_0^j\right)
e^{-T/\tau_{\perp}}\, ,\\
&\langle \Lambda_{zj}\rangle|_{\psi,T}=\cos\Theta_0^j\,
e^{-T/\tau_{\|}}\, .
\end{align}
\end{subequations}
The parameters $\tau_{\perp}$ and $\tau_{\|}$ have been
introduced
to characterize relaxation rates of the electron polarization
due to
the finite width of the ring,
\begin{align}
&\frac{1}{\tau_{\perp}}\equiv\frac{1+\zeta^2}{4} B_{ss}\,
,\label{1_over_tau_perp}\\
&\frac{1}{\tau_{\|}}\equiv2\gamma^2 B_{ss}\,  .\nonumber
\end{align}
As follows from Eq.~(\ref{Lambda_av}), perpendicular to the
$\widetilde{OZ'}$-axis components of $\bm{\Lambda}_j$ 
decay with the rate
$1/\tau_{\perp}$, while the decay rate of parallel components
is $1/\tau_{\|}$. We recall that in the rotating system the
direction
of the $\widetilde{OZ'}$-axis is determined by the vector
$\bm{\theta}$, see (\ref{vector_theta}). At $t=0$ the rotating
system coincides with the original one. Hence, the electron
polarization in the ring relaxes with the rate
$\tau_{\perp}$($\tau_{\|}$), if the polarization of the left
reservoir is perpendicular (parallel) to $\bm{\theta}$. Besides
relaxation associated with finiteness of the ring width, there
is an
additional relaxation channel due to the magnetic field, that
will
be discussed below.

To complete calculation of the spin conductance given by Eqs.
(\ref{gij_av_1}) and (\ref{K_ij_scalar}), the scalar products
$\bm{\Lambda}^0_i\cdot\langle \bm{\Lambda}_j\rangle|_{\psi,T} $, where
$\bm{\Lambda}^0_i$ are given by Eq. (\ref{Lambda_0_i}), 
must be averaged over
$\psi$ and $T$. Averaging over $\psi$ is performed 
with the use of
(\ref{P_W_T}), by substituting $w=-\psi/2\pi \zeta $. As for the distribution
over $T$, in the case of classically chaotic systems one should
use
the exponential function \cite{JBS_Blum_Smil}$\ {\cal
P}(T)$=$\tau^{-1}\exp\left\{-(T-T_0)/\tau\right\}$, where $\tau$
is the mean escape time of a particle and $T_0$ is the shortest
trajectory duration. The results of calculation for $g_{yy}$,
as well as for polarization rotation angles in a magnetic field
are shown in Fig.~\ref{fig3} and Fig.~\ref{fig6} respectively.
It is
seen from these plots that the AC oscillations magnitude and
spin
rotation angle strongly
depend on the parameter $T_1/\tau+T_1/\tau_{\perp}$, which
controls
the trajectory winding number during the particle spin lifetime.
If this parameter is small, $w$ is large and AC oscillations are
strong. In this regime one can write simple analytic expressions for
tensor components of the spin conductance:
\begin{subequations}
\label{spin_cond_result}
\begin{align}
&\langle g_{xx}\rangle =-g_{\,0}\frac{1}{\zeta^2}\left\{Q+4\gamma^2
M\right\}\, ,\\
&\langle g_{yy} \rangle=-g_{\,0}\,Q\, ,\\
&\langle g_{zz} \rangle=g_{\,0}\frac{1}{\zeta^2}\left\{4\gamma^2Q+M\right\}\,
,\\
&\langle g_{xy} \rangle=-\langle g_{yx}\rangle=g_{\,0}\frac{1}{\zeta}R\, ,\\
&\langle g_{xz} \rangle=-\langle g_{zx} \rangle=g_{\,0}\frac{2\gamma}{\zeta^2}\left\{Q-M\right\}\,
,\\
&\langle g_{yz} \rangle=\langle g_{zy} \rangle=g_{\,0}\frac{2\gamma}{\zeta}R\,
,\label{g_zy}\end{align}
\end{subequations}
where $Q, R$ and $M$ are given by
\begin{subequations}
\label{Q_R_M}
\begin{align}
Q&=\cos\pi\zeta\frac{1+\frac{\tau}{\tau_{\perp}}+\frac{2\tau}{T_1}\sin^2\pi\zeta}
{\left(1+\frac{\tau}{\tau_{\perp}}+\frac{2\tau}{T_1}\sin^2\pi\zeta\right)^2
+\left(\frac{\omega_H\tau}{\zeta}\right)^2}\, ,
\\ R&=\cos\pi\zeta\frac{\omega_H\tau/\zeta}
{\left(1+\frac{\tau}{\tau_{\perp}}+\frac{2\tau}{T_1}\sin^2\pi\zeta\right)^2
+\left(\frac{\omega_H\tau}{\zeta}\right)^2}\, ,\label{R}\\
M&=\frac{1}{1+\frac{\tau}{\tau_{\|}}}\, .
\end{align}
\end{subequations}
Note, that expressions (\ref{spin_cond_result}) were obtained
under
an assumption that the magnetic field is not too strong, so
that $\omega_H T_1 /\zeta \ll \sqrt{6\, T_1\, (1/\tau +
1/\tau_{\perp})}$, while
$T_1 /2\, T_0 \gg 1$.

\section{\label{Effects}Results and Discussion}
We start our discussion from the analysis of the finite width
effects. For simplicity, we will consider the
$\langle g_{yy}\rangle$
component of the spin conductance matrix in the regime of large
winding numbers when analytic expressions
(\ref{spin_cond_result})
are valid. In the absence of the magnetic field the spin
conductance
is given by
\begin{equation}
\label{g_yy_H_0}
\langle g_{yy}\rangle=-g_0\cos\pi\zeta
\frac{\frac{1}{\tau}}{\frac{2}{T_1}\sin^2\pi\zeta+\frac{1}{\tau}+\frac{1}{\tau_{\perp}}}\,
.
\end{equation}

\begin{figure}
\includegraphics[width=247pt]{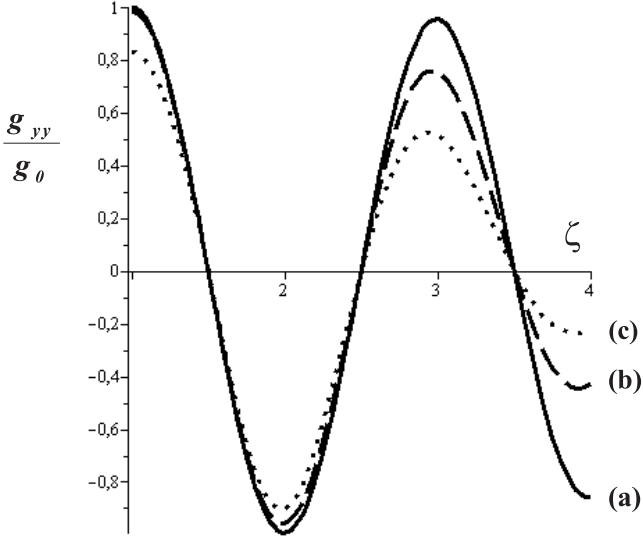}
\caption{Attenuation of spin current oscillations
due to finite widths of the  ring and leads, at a zero
magnetic field.
$g_{yy}$ is the $y$-component of the spin conductance
along the $y$-axis, see (\ref{J}) in the text. $g_0$ is the 
transmitted
flux of electrons per unit energy interval, per one spin 
projection of the incident electrons.
The variable 
$\zeta$ is expressed in terms of the spin-orbit length $L_{SO}$ 
as
$\zeta=\sqrt{1+4(a/L_{SO})^2}$.
Three curves  correspond
to parameter sets \\
(a) $ a = 1\:\mu \,,\,\,d =150\:nm \,\,,\,\,W =150\:nm
\,\,,\,\,q = 1/2\,$;\\
(b) $ a = 1,5\:\mu \,,\,\,d =300\:nm \,\,,\,\,W
=100\:nm
\,\,,\,\,q = 1/5\,$;\\
(c) $ a = 1,5\:\mu \,,\,\,d =300\:nm \,\,,\,\,W
=30\:nm\,\,,\,\,q = 1/5\,$.}
\label{fig3}
\end{figure}
The denominator in (\ref{g_yy_H_0}) gives rise to a set of
peaks with maxima at $\zeta=\zeta_m\equiv m$, where $m$ is
integer. It
is
easily seen that for $T_1^{-1}\gg\tau^{-1}+\tau_{\perp}^{-1}$
the peak's broadening $\Delta_g \ll 1$.
Due to this inequality Eq. (\ref{g_yy_H_0})
can be written in the vicinity of peaks in a more simple form:
\begin{equation}
\label{g_yy_peak}
\langle g_{yy}\rangle =(-1)^{m+1} g_0
\frac{\frac{T_1}{2\pi^2}\frac{1}{\tau}}
{\left(\Delta\zeta\right)^2+\frac{T_1}{2\pi^2}\left(
\frac{1}{\tau}+\frac{1}{\tau_{\perp}}\right)}\, ,
\end{equation}
where $\Delta \zeta=\zeta-\zeta_m$. From Eq. (\ref{g_yy_peak})
$\Delta_g$ is expressed as
\begin{equation}
\label{Delta_g} \Delta_g=\sqrt{\frac{2
T_1}{\pi^2}\left(\frac{1}{\tau}+\frac{1}{\tau_{\perp}}\right)}\,.
\end{equation}
Note, that as follows from (\ref{tau_perp_expr}),
$\tau_{\perp}$ depends on $\zeta$. Hence, $\Delta_g$ depends on
the resonance
$\zeta_m$ position.

In the case of a long particle lifetime $\tau \gg \tau_{\perp}$
one
obtains from Eq. (\ref{tau_perp_expr})
\begin{equation}
\label{Delta_g2}
\Delta_g=\gamma_m^2\frac{d}{a}\sqrt{\frac{8(1+\zeta^2_m)}{3\zeta^2_m}}\,
.
\end{equation}
For example, the broadening of the third peak ($\zeta_m=3,
\gamma_m=\sqrt{2}$) is $\Delta_g\simeq 3.4 d/a$. It is a quite
noticeable value for a typical ratio $d/a \sim 0.1$

\begin{figure}
\includegraphics[width=247pt]{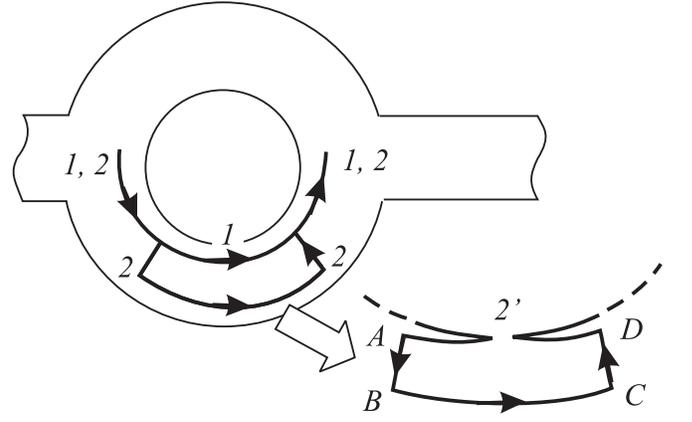}
\caption{A schematic picture explaining the effect of finite
width. Since the element of the trajectory passed forward and
backward does not give a contribution to the evolution
operator,one can replace the trajectory $2$ by $2'$. After that
it is
obvious that the difference between $1$ and $2$ is that the
latter
contains the loop $ABCD$ passed counterclockwise. }
\label{fig4}\end{figure}

So, the first obvious effect of the finite width is the
broadening
of the spin current oscillation peaks. The physical origin of
this
effect is the increased relaxation rate of the spin
polarization.
This relaxation is caused by incoherent superposition of
polarizations coming from the trajectories encircling slightly
different
areas in a ring of finite width. The situation is elucidated in
Fig.~\ref{fig4}. This picture shows that the finite width
results in
adding random loops breaking the coherency of the trajectories.

In addition to the broadening, the increased relaxation rate
leads,
evidently, to a reduction of the peak intensity. This is
explicitly
given by
\begin{equation}
\label{g_magnitude}
\left|\left< g_{yy}\right>_{\zeta=\zeta_m}\right|=\frac{g_0}
{1+\frac{\tau}{\tau_{\perp}}}\,
\end{equation}
which is the spin current magnitude exactly at maxima (minima).
From
Appendix~\ref{char_times} the ratio of times in the denominator
of
(\ref{g_magnitude}) can be expressed as
\begin{equation}
\label{tau_tau_perp}
\frac{\tau}{\tau_{\perp}}\simeq\frac{1}{6\pi q}
\frac{\gamma^4 (1+\zeta^2)}{\zeta^2}
\frac{a}{W}\left(\frac{d}{a}\right)^4 \ln^2\frac{2a}{d} \, .
\end{equation}
This expression shows that the finite width effect is
suppressed fast with smaller $d/a$.

Now let us focus on magnetic field effects. The first effect is
that
the components $\langle g_{yz}\rangle$, $\langle g_{zy}\rangle$,
$\langle g_{xy}\rangle$, $\langle g_{yx}\rangle$ of the spin
conductance
are no longer zero.  One can verify from Eqs. (\ref{gij_av_1}),
(\ref{Kij}) and (\ref{expr_S}) that they appear because the
reflection symmetry with
respect to the $XOZ$-plane is broken by the magnetic field. The
physical meaning of such nondiagonal components can be explained in
terms of the effective polarization $\bm{P}_{eff}(\bm{e}_j)$ on
the
exit from the ring, see Eq. (\ref{P_eff_full}). For example,
nonzero
$\langle g_{zy}\rangle$ and $\langle g_{xy}\rangle$ are
associated with a rotation of $\bm{P}_{eff}(\bm{e}_y)$ with
respect to the
polarization $\bm{e}_y$ of the left reservoir. It is convenient
to
consider a projection of $\bm{P}_{eff}(\bm{e}_y)$ onto
$YOZ$-plane. Then, the rotation angle $\phi_P$ of this
projection
can be calculated from Eqs. (\ref{spin_cond_result}) and
(\ref{Q_R_M}). For the $m$-th peak this angle is given by
\begin{equation}
\label{tan_phi_P_zeta_m}
\tan
\phi_P=-\frac{2\gamma}{\zeta_m}\frac{\frac{\omega_H}{\zeta_m}}
{\frac{1}{\tau}+\frac{1}{\tau_{\perp}}}\, .
\end{equation}
The nonzero $\langle g_{yz}\rangle$, $\langle g_{yx}\rangle$
components
can be interpreted in a similar way. We note, that due to the
linear
dependence on $\omega_H$, the sign of $\phi_P$ changes together
with
the magnetic field.

\begin{figure}
\includegraphics[width=247pt]{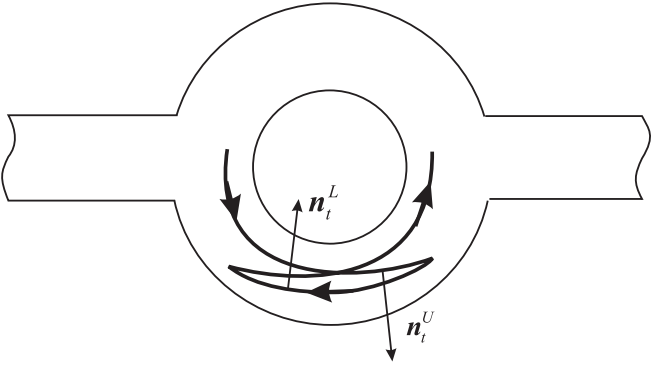}
\caption{Narrow loops of the trajectory which break the
coherence
of the trajectories in the presence of the magnetic field.}
\label{fig5}
\end{figure}

Another effect of the magnetic field is a reduction of the spin
current. Let us consider a trajectory which contains a narrow
loop, see Fig.~\ref{fig5}.
If the magnetic field is ignored and only the SOI effect is
taken
into account, after passing the loop the polarization $\bm{P}$
does
not change. That is because on the upper and the lower parts of
the
loop $\bm{P}$ rotates in opposite directions, according to
opposite
directions $\bm{n}_t^U$ and $\bm{n}_t^L$ of the SOI fields, see
Eq.
(\ref{expr_S}). A magnetic field, however, causes rotations of
$\bm{P}$ in the same directions. Hence, evolutions of $\bm{P}$
along
trajectories with and without the loop becomes different. This
introduces an additional decoherence leading to the spin
current reduction and broadening of its oscillation peaks. Using
(\ref{spin_cond_result}) and (\ref{Q_R_M}) one can derive the
following expression for the magnitude of the effective
polarization
exactly at maxima (minima)
\begin{align}
\label{P_eff_H_zeta_m} &\left.
P_{eff}(\bm{e}_y)\right|_{\zeta=\zeta_m}
\nonumber\\
&=g_0^{-1}\sqrt{\left|\langle g_{yy}\rangle_{\zeta=\zeta_m}\right|^2
+\left|\langle g_{zy}\rangle_{\zeta=\zeta_m}\right|^2
+\left|\langle g_{xy}\rangle_{\zeta=\zeta_m}\right|^2}\nonumber\\
&\simeq\left[1+\left(\frac{\omega_H\tau}{\zeta_m}\right)^2\right]^{-1/2}\,
,
\end{align}
provided that the ring is narrow enough,
$\tau/\tau_{\perp} \ll \omega_H\tau /\zeta_m$.

Let us consider a dependence of the spin conductance 
on $\zeta$ in the presence of the magnetic field.
In the the vicinity of the $m$-th peak, 
instead of (\ref{g_yy_peak}) wehave
\begin{equation}
\label{g_yy_H} \langle g_{yy}\rangle=g_0 (-1)^{m+1}
\frac{\frac{1}{\tau}\left[\frac{1}{\tau}+\frac{1}{\tau_{\perp}}+
\frac{2\pi^2}{T_1}(\Delta
\zeta)^2\right]}{\left[\frac{1}{\tau}+\frac{1}{\tau_{\perp}}+
\frac{2\pi^2}{T_1}(\Delta
\zeta)^2\right]^2+\left(\frac{\omega_H}{\zeta_m}\right)^2}\,.
\end{equation}
\begin{figure}
\includegraphics[width=247pt]{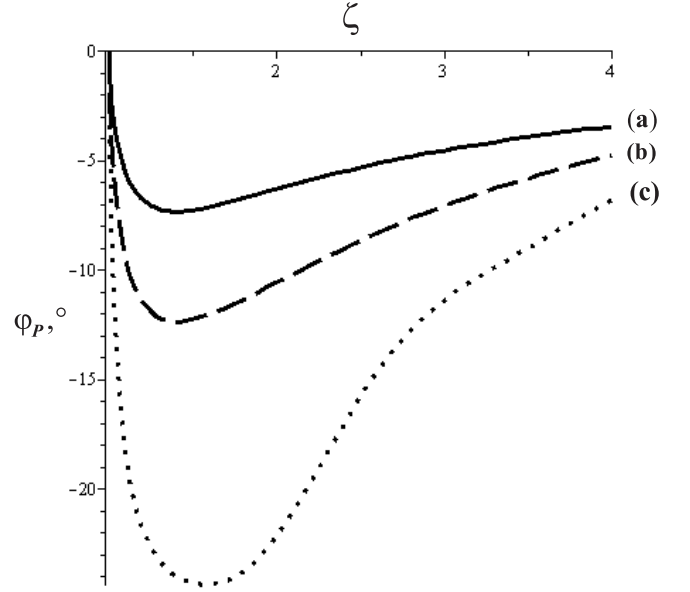}
\caption{Rotation of the effective polarization vector
$\bm{P}_{eff}(\bm{e}_y)$ on the exit from the ring in the
presence of the magnetic field. $\phi_P$ is the angle between
the initial polarization $\bm{e}_y$ of the electrons in the left
reservoir and projection of $\bm{P}_{eff}(\bm{e}_y)$ onto
the
$YOZ$
plane. The magnetic field strength is 100G.
Curves (a), (b) and (c) correspond
to the same parameter sets as in Fig.~{\ref{fig3}}.}
\label{fig6}
\end{figure}
As can be seen from this equation, the relaxation mechanism
associated with the magnetic field gives rise to an additional
broadening of spin current peaks. Their width can be evaluated
from
Eq. (\ref{g_yy_H}) as
\begin{equation}
\label{Delta_g_H} \Delta_g=\sqrt{\frac{2T_1}{\pi^2}}
\left[\left(\frac{1}{\tau}+\frac{1}{\tau_{\perp}}\right)^2
+\left(\frac{\omega_H}{\zeta_m}\right)^2\right]^{1/4}.
\end{equation}

The discussed above effects are determined by characteristic
times
$\tau$, $T_1$ and $\tau_{\perp}$. These times have been
evaluated in
Appendix C using a simple model of scattering from a bumpy ring
boundary. 
Fig.~\ref{fig3} and Fig.~\ref{fig6} demonstrate the effects of
the ring width and magnetic field on behavior
of the spin conductance as a function of 
$\zeta=\sqrt{1+4(a/L_{so})^2}$. 
We took $k_F\simeq 2.5\times 10^6\: cm^{-1}$ and
$L_{so}$
varying in a wide range. In InAs based quantum wells
SOI can be quite strong with $L_{so}$ being as small, 
as $\sim 100$~nm\cite{Grundler,Nitta}. 
For $a=1\mu$
this gives $\zeta=20$. 
Curve (a)
in these figures corresponds to 
the escape time much shorter than
$T_1$ and $\tau_{\perp}$. The winding number is not 
large and
AC resonances are broad. 
There are no 
noticeable effects associated with the finite ring width.  
Also, the magnetic field effect is relatively weak. 
The width effects are seen 
on the curve (b),
in  Fig.~\ref{fig3}. A reduction of
the oscillation amplitude seen in the figure is in a 
qualitative agreement with Eq.~(\ref{g_magnitude}), 
although for considered parameters this
equation can not be fully applied, because the winding number is not
large enough. The winding number is larger for the third set of
parameters, (c). The finite width effect becomes stronger
, leading to a 
faster decreasing of the oscillation amplitude. 
Also stronger is the magnetic field effect on
polarization rotation in the $xy$-plane (see Fig. \ref{fig6}).
Strictly speaking, our semiclassical theory can not be applied
to
this case, because the number of propagating channels in leads
is
not large. We, nevertheless show this result in order to
demonstrate
a trend: for reasonable ring sizes the regime of large windings
with
sharp AC resonances can be achieved only at small lead widths,
or by
means of barriers between leads and the ring, resulting in the
long~$\tau$.

We note that the magnetic field effect on polarization rotation
is
rather noticeable even at relatively weak 100 Gauss magnetic
fields,
as can be seen at Fig. \ref{fig6}. For example, for $\zeta=1.6$
and parameters
(c), the rotation angle can be as large as
$24^{\circ}$.

\begin{acknowledgments}
This work has been supported by RFBR Grant No. 060216699.
V.V.S. also acknowledges support from RFBR Grant No. 09-02-01235.
\end{acknowledgments}

\appendix
\section{\label{Landauer}Landauer formula for the spin current}

Since the spin current density is an additive one-particle
dynamical observable, its average value at the point ${\bf R}$
and
$t=0$ is given by
\begin{equation}
\label{sp_cu expr} \langle J_{lj}({\bf R},t=0) \rangle = {\rm Tr}
\left\{
\hat{f}_1(t=0)\hat{J}_1^{lj}({\bf R}) \right\}
\end{equation}
Here the one-particle distribution function $\hat{f}_1$
describes
the open system consisting of the leads and the ring. The
one-particle operator
\begin{equation}
\label{def_sp_cu} \hat{J}_1^{lj}({\bf R})=
\frac{\hat{v}_l \hat{{\cal P}}_{j}({\bf R})+\hat{{\cal
P}}_{j}({\bf R})\hat{v}_l} {2}
\end{equation}
represents $l$-th component of the current density with spins
polarized
along $j$-th coordinate axis. $\hat{v}_l$ is $l$-th component of the
electron velocity operator. Since we calculate the spin current
in
the asymptotic region of the right lead, where the magnetic
field
and SOI are zero, the operator $\hat{v}_i$ may be written simply as
$\hat{p}_i/m^*$, where $\hat{\mathbf{p}}$ is
the electron momentum operator. The polarization density
$\hat{{\cal
P}}_{j}({\bf R})$ is defined by
\begin{equation}
\label{def_sp_den} \hat{{\cal P}}_{j}({\bf R})=
                       \hat{\rho}_{\bf R} \, \hat{\sigma_j}\, ,
\end{equation}
where $\hat{\rho}_{\bf R}$ is the density operator. In the the
coordinate representation the latter is given by
\begin{equation}
\label{density} \hat{\rho}_{\bf R}=
                                  \delta({\bf r}-{\bf R})\, .
\end{equation}
It should be noted that Eq. (\ref{sp_cu expr}) represents a
polarization current density, rather then the spin current
density,
which is twice smaller. For convenience we will use, however,
the
latter name.

For noninteracting electrons the evolution of the distribution
function is described by the equation
\begin{equation}
\label{f_1_evol_eq} i \hbar \frac{\partial \hat{f}_1}{\partial
t}=
                    [H,\hat{f}_1]        \,
\end{equation}
where $H$ is the one-particle Hamiltonian for the system
"leads+ring". A formal solution of Eq.(\ref{f_1_evol_eq}) may
be written in the form
\begin{gather}
\hat{f}_1(t)=\hat{U}(t,t_0)\hat{f}_1(t_0)\hat{U}^+(t,t_0)\,
,\label{f_1_evol_expr}\\
\hat{U}(t,t_0)=e^{-i\hat{H}(t-t_0)/\hbar}\, .\label{U}
\end{gather}
Further, we take into account that the system under
consideration
has an asymptotic region where an electron is effectively
decoupled
from the ring. In this case the methods of the scattering theory may
be applied directly, without adiabatic switching off the
scattering
potential at $t=\pm \infty$. First, let us write down
$\hat{f}_1(0)$
in the form
\begin{multline}
\label{f_1_lim} \hat{f}_1(0)=\lim_{T\rightarrow + \infty}
\hat{U}(0,-T)\hat{U}_0^+(0,-T)\hat{U}_0(0,-T)\\
\times\hat{f}_1(-T)\hat{U}_0^+(0,-T)\hat{U}_0(0,-T)\hat{U}^+(0,-T)\,
,
\end{multline}
where the unperturbed evolution operator $\hat{U}_0$ is
obtained from (\ref{U}) by replacing $\hat{H}$ with
the "unperturbed" Hamiltonian $\hat{H}_0$. The latter is
obtained by removing the ring and
elongating the leads to meet each other. In Eq. (\ref{f_1_lim})
one
easily recognizes the familiar M\"{o}ller operator $\Omega_+$ of
 the scattering theory \cite{Taylor},
\begin{equation}
\label{Omega_plus_def} \Omega_+=\lim_{T\rightarrow + \infty}
\hat{U}(0,-T)\hat{U}_0^+(0,-T) \, .
\end{equation}
This operator maps the wave function $\left|\psi_{in}\right>$
describing a particle state at $t=0$ in the absence of the ring
onto
the actual state $\left|\psi(t=0)\right>$ :
\begin{equation}
\label{psi_Omega_plus_psi_in} \left|\psi(t=0)\right>=\Omega_+
\left|\psi_{in}\right>\, .
\end{equation}
From (\ref{Omega_plus_def}) and (\ref{f_1_lim}) we obtain
\begin{equation}
\label{f_1_Omega_plus} \hat{f}_1(0)=\Omega_+ \hat{f}_1^{in}
\left(\Omega_+ \right)^+ \, ,
\end{equation}
where, by analogy with $\left|\psi_{in}\right>$, the function
\begin{equation}
\label{f_1_in} \hat{f}_1^{in}=\lim_{T\rightarrow
+\infty}\hat{U}_0(0,-T)\hat{f}_1(-T)\hat{U}_0^+(0,-T)
\end{equation}
can be interpreted as a distribution function of the system at
$t=0$
in the absence of the ring. The trace in (\ref{sp_cu expr}) can
now
be rewritten as
\begin{multline}
\label{sp_cu_sum_over_i} \langle J_{lj}({\bf R},t=0) \rangle =
\sum_{i_1, i_2} \left< i_1 \right|\hat{f}_1^{in}\left|i_2\right>\\
\times \left< i_2
\right|\left(\Omega_+\right)^+\hat{J}_1^{lj}\left({\bf
R}\right)\Omega_+\left|i_1\right> \, .
\end{multline}
Since the unperturbed problem does not involve SO interaction,
a convenient choice of the basis vectors $\left|i\right>$ in
(\ref{sp_cu_sum_over_i}) is
\begin{equation}
\label{i}
\left|i\right>=\left|b\right>\otimes\left|\alpha\right>\,
,
\end{equation}
where the eigenvector $\left|b\right>$ corresponding to the
unperturbed Hamiltonian describes the electron orbital motion
and
$\left|\alpha\right>$ is the eigenvector of $\hat{\sigma}_z$
corresponding to its eigenvalue $\alpha$. Further, the slab
geometry
of the unperturbed problem suggests that $\left|b\right>$ is
taken
in the form
\begin{equation}
\label{b}
\left|b\right>=\left|k\right>\otimes\left|p\right>\otimes\left|m\right>\,
,
\end{equation}
with the eigenvectors
$\left|k\right>,\left|p\right>,\left|m\right>$
describing a particle motion along $OX,OY,OZ$ axes in the
absence of
the ring. Hence, the corresponding wave functions are
\begin{gather}\label{v_p}
w_k(x)=\left< x \right|\left. k \right>= \frac{1}{\sqrt{L}}
e^{ikx} \, ,\\
v_p(y)= \left< y \right|\left. p \right>=\sqrt{\frac{2}{L_y}}\,
sin{(k_y y)} \\
k_y=\frac{\pi}{L_y}p \, ,\,\, (p=1,2,...)
\end{gather}
and similarly for the wave function 
$u_m(z),\, m=1,2,...$, 
in z-direction. We took periodic
boundary conditions in x-direction, where $L$ is the total
length of
the system. At the slab interfaces the wave functions $v_p(y)$
and
$u_m(z)$ satisfy the hard wall boundary conditions.

Unit vectors parallel to polarizations of the left and right
reservoirs will be denoted as $\bm{\nu}^L$ and $\bm{\nu}^R$,
respectively. Accordingly, we define the operators
$\hat{\sigma}_{\nu^{L,R}} \equiv \sum_i \hat{\sigma}_i
\nu^{L,R}_i$
with eigenvectors $\left|\nu^{L,R}\sigma\right>$
corresponding to polarization projections
$\sigma = \uparrow,\downarrow$ onto
$\bm{\nu}^L$ and $\bm{\nu}^R$. Since particles with different
spins
are distributed in reservoirs according to their respective
Fermi
distributions, the magnitudes of the reservoirs polarizations
are
determined by the differences
$\delta\mu_{L,R}=\mu_{\nu^{L,R}\uparrow}-\mu_{\nu^{L,R}\downarrow}$
of chemical potentials of spin up and spin down (relative to
$\bm{\nu}^{L,R}$)
electron gas components.  Therefore, assuming that
the unperturbed distributions of particles moving to the right
($k>0$)
and to the left ($k<0$) are given by the Fermi distributions in
the
left and right reservoirs, respectively, we can write
\begin{multline}
\label{f_1_in_expr} \hat{f}_1^{in} = \sum_{b, \sigma}\Theta
\left(k\right)n_{b,\nu^L \sigma} \left(\left|b\right>\left<
b\right|\right)\otimes
\left(\left|\nu^L\sigma\right>\left<
\nu^L\sigma\right|\right)\\+ \sum_{b, \sigma}\Theta
\left(-k\right)n_{b,\nu^R \sigma}
\left(\left|b\right>\left< b\right|\right)\otimes
\left(\left|\nu^R\sigma\right>\left< \nu^R\sigma\right|\right)\,,
\end{multline}
where $\Theta (\cdot)$ is the Heaviside step function and
\begin{equation}
\label{n_b_nu_sigma_L} n_{b,\nu^{L} \sigma}=\Theta
\left(\mu_{\nu^{L}\sigma}-E\right)\,,\,\,n_{b,\nu^{R}
\sigma}=\Theta\left(\mu_{\nu^{R}\sigma}-E\right)
\end{equation}
are the Fermi distributions in
the left and right reservoirs for particles with the energy
$E$. Note, that Eq. (\ref{f_1_in_expr}) was written under the
assumption
that contacts between reservoirs and leads are adiabatic (no
scattering from the contacts).

We assume for the average chemical potentials
$\mu_{L}=\mu_{R}$, where
$\mu_{L/R}=(\mu_{\nu^{L/R}\uparrow}+\mu_{\nu^{L/R}\downarrow})/2$.
So, the chemical
potentials of unpolarized reservoirs coincide. Denoting them
$\mu_U$
we write the distribution function corresponding to the
unpolarized
reservoirs in the form
\begin{equation}
\label{f_1_U_def} \hat{f}_1^{U} = \sum_{b} n^U_{b}
\left(\left|b\right>\left< b\right|\right)\otimes \sigma_0,
\end{equation}
where $n^U_b=\Theta(\mu_U-E)$. Evidently, $\hat{f}_1^{U}$ does
not
give any contribution to $\langle J_{lj}({\bf R},t=0) \rangle$.
Therefore, it is convenient to subtract this function from
$\hat{f}^{in}_1$:
\begin{gather}
\label{f_1_U_sep}
\hat{f}_1^{in} = \hat{f}_1^{U}+\delta \hat{f}_1^{L}+\delta
\hat{f}_1^{R}\\
\delta\hat{f}_1^{L} =\sum_{b, \sigma} \Theta \left(k\right)
\delta
n_{b,\nu^{L} \sigma} \left(\left|b\right>\left<
b\right|\right)\otimes
\left(\left|\nu^{L}\sigma\right>\left<
\nu\sigma^{L}\right|\right)\label{delta_f_1_L_def}\\
\delta\hat{f}_1^{R} =\sum_{b, \sigma} \Theta \left(-k\right)
\delta
n_{b,\nu^{R} \sigma} \left(\left|b\right>\left<
b\right|\right)\otimes
\left(\left|\nu^{R}\sigma\right>\left<
\nu^{R}\sigma\right|\right)\label{delta_f_1_R_def}\\
\delta n_{b,\nu^{L,R}\sigma}=n_{b,\nu^{L,R}\sigma} - n_{b}^U
\label{delta_n_def}
\end{gather}
Denoting corresponding contributions of $\delta\hat{f}_1^{L}$
and
$\delta\hat{f}_1^{R}$ to the spin current as $\langle
J_{lj}({\bf R},t=0) \rangle_{L}$ and $\langle J_{lj}({\bf R},t=0)
\rangle_{R}$,
we arrive at
\begin{equation}
\label{J_J_L_J_R} \langle J_{lj}({\bf R},t=0) \rangle= \langle
J_{lj}({\bf R},t=0) \rangle_{L}+ \langle J_{lj}({\bf R},t=0)
\rangle_{R}\, .
\end{equation}
The projectors $\left|\nu^{L}\sigma\right>\left<
\nu^{L}\sigma\right|$ and $\left|\nu^{R}\sigma\right>\left<
\nu^{R}\sigma\right|$ in (\ref{delta_f_1_L_def}) and
(\ref{delta_f_1_R_def}) can be expressed  in terms of the Pauli
matrices $\hat{\sigma}_{\nu^{L,R}}$ and the unit matrix
$\hat{\sigma}_0$ using easily verified relations
\begin{eqnarray*}
\hat{\sigma}_{\nu}= \left(\left|\nu\uparrow\right>\left<
\nu\uparrow\right|\right)-
                    \left(\left|\nu\downarrow\right>\left<
                    \nu\downarrow\right|\right)\\
\hat{\sigma}_{0}=
\left(\left|\nu\uparrow\right>\left<\nu\uparrow\right|\right)+
                    \left(\left|\nu\downarrow\right>\left<
                    \nu\downarrow\right|\right)
\end{eqnarray*}
Straightforward calculations then give
\begin{multline}
\label{sp_cu_sum_k_dE} \langle J_{lj}({\bf R},t=0) \rangle_{L,R}
=\hspace{-7mm} \sum_{ \substack{
b, \alpha_1, \alpha_2\\
\mu_U-\frac{\delta \mu_{L,R}}{2} <E <\mu_U + \frac{\delta
\mu_{L,R}}{2} } }
\hspace{-7mm}\Theta(\pm
k)\frac{\sigma_{\nu^{L,R}}^{\alpha_1\alpha_2}}{2}\\
\times\left< b \alpha_2 \right|
\left(\Omega_+\right)^+\hat{J}_1^{lj}\left({\bf
R}\right)\Omega_+
\left|b\alpha_1\right> \, ,
\end{multline}
where $\left| b \alpha_{1,2}\right> \equiv \left|b\right>
\otimes
\left| \alpha_{1,2}\right> $,
and upper (lower) sign in the argument of
$\Theta$-function corresponds to the index "L" ("R"). The leads
are
assumed to be thin enough in z-direction, so that only the
levels
$\left|b\right>=\left|k p m\right>$ with $m=1$ are occupied and
contribute to the sum in Eq. (\ref{sp_cu_sum_k_dE}). For
simplicity, we denote
\begin{equation}
\label{kp_def} \left|kp\right>=\left|kp,m=1\right> \, ,
\end{equation}
It is convenient to change in (\ref{sp_cu_sum_k_dE}) the
summation
over $k$ by integration over $E$. To do this, we introduce
the vectors
\begin{gather}
\left|Ep\right>^{(\pm)}=
\sqrt{\nu_p(E)}\left|kp\right> \label{Ep_def}\\
\nu_p(E)=\frac{L}{2\pi \hbar \, v_p(E)}\, , \nonumber
\end{gather}
where $\pm$ signs relate to $k>0$ and $k<0$. $\nu_p(E)$ is the
one-dimensional density of states in the $p$-th channel,
and $v_p(E)=\sqrt{2(E-E_p)/m^*}$ is
the electron velocity in $p$-th channel characterized by the
kinetic
energy in $y$-direction $E_p$. Using definitions (\ref{Ep_def})
one
can write
\begin{multline}
\label{b_Ep_rel} \sum_{
\substack{b\\
\mu_U-\frac{\delta \mu_{L,R}}{2} <E <\mu_U + \frac{\delta
\mu_{L,R}}{2} } } \hspace{-7mm} \Theta(\pm k)
\left< b \right|      \cdot     \left|b\right>\\
=\sum_p \int_{\mu_U-\frac{\delta \mu_{L,R}}{2}}^{\mu_U +
\frac{\delta \mu_{L,R}}{2}}dE \hspace{2mm}
{^{(\pm)}}\hspace{-1mm}\left< Ep \right|      \cdot
\left|Ep\right>^{(\pm)} \, ,
\end{multline}
Further, in the limit $L\rightarrow +\infty$ considering $E$ as
a
continuous variable one gets the normalization condition
\begin{equation}
\label{Ep_normalization} {^{(\pm)}}\hspace{-1mm}\left< E'p'
\right|\left. Ep\right>^{(\pm)} = \delta_{p'p}\delta(E'-E)\, .
\end{equation}
Using this condition and substituting (\ref{b_Ep_rel}) into
Eq.~(\ref{sp_cu_sum_k_dE}) we find
\begin{multline}
\label{sp_cu_int_dE} \langle J_{lj}({\bf R},t=0) \rangle_{L,R}
=\sum_{p, \alpha_1, \alpha_2} \int_{\mu_U-\frac{\delta
\mu_{L,R}}{2}}^{\mu_U + \frac{\delta \mu_{L,R}}{2}} dE \,
\frac{\sigma_{\nu^{L,R}}^{\alpha_1\alpha_2}}{2}\\
\times{^{(\pm)}}\hspace{-1mm}\left< Ep\,\alpha_2 \right|
\left(\Omega_+\right)^+\hat{J}_1^{lj}\left({\bf
R}\right)\Omega_+
\left|Ep\,\alpha_1\right>^{(\pm)} \, ,
\end{multline}
($\left|Ep\, \alpha_{1,2}\right>^{(\pm)}\equiv
\left|Ep\right>^{(\pm)} \otimes \left|\alpha_{1,2}\right>$). In
its turn, the total spin current through the cross section of 
the right lead is given by
\begin{equation*}
\langle J_j(X,t=0) \rangle_{L,R}=\int dY dZ \langle J_{xj}({\bf
R},t=0) \rangle_{L,R}\, .
\end{equation*}
Using Eq. (\ref{sp_cu_int_dE})  we obtain
\begin{multline}
\label{total_sp_cu_int_dE}
\langle J_j(X,t=0) \rangle_{L,R}\\
=\sum_{p, \alpha_1, \alpha_2} \int_{\mu_U-\frac{\delta
\mu_{L,R}}{2}}^{\mu_U + \frac{\delta \mu_{L,R}}{2}} dE \,
\frac{\sigma_{\nu^{L,R}}^{\alpha_1\alpha_2}}{2}\int dY dZ\\
\times{^{(\pm)}}\hspace{-1mm}\left< Ep\,\alpha_2,+ \right|
\hat{J}_1^{xj}\left({\bf R}\right)
\left|Ep\,\alpha_1,+\right>^{(\pm)} \, ,
\end{multline}
where
\begin{equation*}
\left|Ep\,\alpha_s,+\right>^{(\pm)}=\Omega_+\left|Ep\,\alpha_s\right>^{(\pm)}\,,\,\,
s=1,2\, .
\end{equation*}
The vectors $\left|Ep\,\alpha_s,+\right>^{(\pm)}$ are known
\cite{Taylor} as the scattering states associated with the
"in" asymptotes ("incident waves")
$\left|Ep\,\alpha_s\right>^{(\pm)}$.
Since the point $\bf R$ in the r.h.s of
(\ref{total_sp_cu_int_dE}) is located
in the asymptotic region of the right lead,
only the asymptotic behavior of the wave functions
$\phi^{(\pm)}_{Ep\, \alpha_2,+} ({\bf R}\alpha)\equiv
\left<{\bf R}\alpha\right|\left.Ep\, \alpha_2,+\right>^{(\pm)}$
and
$\phi^{(\pm)}_{Ep\, \alpha_1, +} ({\bf R}\beta)$
affects the calculation of the matrix elements in
(\ref{total_sp_cu_int_dE}).
Thus, we may write $\phi^{(+)}_{Ep\, \alpha_2,+} ({\bf
R}\alpha)$ and
$\phi^{(+)}_{Ep\, \alpha_1,+} ({\bf R}\alpha)$ as the sum of
transmitted waves while
$\phi^{(-)}_{Ep\, \alpha_2,+} ({\bf R}\alpha)$ and
$\phi^{(-)}_{Ep\, \alpha_1,+} ({\bf R}\alpha)$
as the sum of incident and reflected waves,
\begin{subequations}
\label{phi_plus_minus_Epalpha_expr}
\begin{align}
\phi^{(+)}_{Ep\, \alpha_2,+}({\bf R}\alpha)&=
\sum_{p''\alpha''}t_{p'' p}^{\alpha'' \alpha_2}
(E)\phi^{(+)}_{Ep''\alpha''}({\bf R}\alpha) \\
\phi^{(-)}_{Ep\, \alpha_2,+}({\bf R}\alpha) &= \phi^{(-)}_{Ep\,
\alpha_2}({\bf R}\alpha) \nonumber\\
&+ \sum_{p''\alpha''}r_{p'' p}^{\alpha'' \alpha_2}
(E)\phi^{(+)}_{Ep''\alpha''}({\bf R}\alpha),
\end{align}
\end{subequations}
where $t_{p'' p}^{\alpha'' \alpha_2}(E)$ and 
$r_{p'' p}^{\alpha'' \alpha_2}(E)$
denote transmission and reflection amplitudes, respectively.
According to (\ref{Ep_def}) and (\ref{kp_def})
\begin{equation}
\label{phi_Epalpha_ksi} \phi^{(\pm)}_{Ep''\,\alpha''}({\bf
R}\alpha)=\xi^{(\pm)}_{p''E}(X)u_{p''}(Y)v_1(Z)\chi_{\alpha''}(\alpha)\,
,
\end{equation}
where
\begin{equation}
\label{xi} \xi^{(\pm)}_{p''E}(X)=\sqrt{\nu_{p''}(E)}w_{\pm
k}(X)=\frac{1}{\sqrt{2 \pi \hbar v_{p''}(E)}}\, e^{\pm ikX},
\end{equation}
 and
$\chi_{\alpha''}(\alpha)=\left<\alpha\right|\left.\alpha''\right>$.
$k$ in (\ref{xi}) is the positive solution of the equation
$E-E_{p"}=\hbar^2 k^2/2m^*$. Note, that due to our choice of
the prefactor in (\ref{Ep_def}), the functions $\xi_{p''E}(X)$
"carry"
the same flux $(2\pi \hbar)^{-1}$, independent on the 
channel number $p''$. As a result, 
the transmission and reflection amplitudes satisfy
the flux conservation law
\begin{equation*}
\sum_{p"\alpha"}\left\{\left|t_{p"p}^{\alpha"\alpha_2}\right|^2
+\left|r_{p"p}^{\alpha"\alpha_2}\right|^2\right\}=1
\end{equation*}
Using Eq. (\ref{def_sp_cu}) and Eqs.
(\ref{phi_plus_minus_Epalpha_expr}) -- (\ref{phi_Epalpha_ksi}),
we transform Eq. (\ref{total_sp_cu_int_dE}) into

\begin{subequations}
\label{J_j_L_R_t+_sigma_j_t_sigma_nu}
\begin{gather}
\begin{split}
& \langle J_j\left(X,t=0\right)\rangle_L\\
& =\frac{\delta \mu_L}{2 \pi \hbar} 
\sum_{p,p'}
\frac{1}{2}\,{\rm Tr}\left\{\hat{t}_{p'p}^+(\mu_U)\,\hat{\sigma}_{j}\, 
\hat{t}_{p'p}(\mu_U)\,\hat{\sigma}_{\nu^L}\,\right\}
\end{split}
\\
\begin{split}
\langle J_j\left(X,t=0\right)\rangle_R \quad &\\
=\frac{\delta \mu_R}{2 \pi \hbar}
\left[\vphantom{\frac{1}{2}}\right.\sum_{p,
p'}\frac{1}{2} & \,{\rm Tr}\left\{\hat{r}_{p'p}^+(\mu_U)\,\hat{\sigma}_{j}
\,\hat{r}_{p'p}(\mu_U)\,\hat{\sigma}_{\nu^R}\,\right\} \\
& -\left. N_m(\mu_U)
\frac{1}{2}\,
{\rm Tr}\left\{\hat{\sigma}_{\nu^R}\hat{\sigma}_{j}\right\}\right]
\end{split}
\end{gather}
\end{subequations}
where $\hat{t}_{p'p}(\mu_U)$ is the $2\times 2$ matrix composed of
the transmission amplitudes
$t_{p'p}^{\uparrow\uparrow}(\mu_U)$,$\,t_{p'p}^{\uparrow\downarrow}(\mu_U)\,$,
$t_{p'p}^{\downarrow\uparrow}(\mu_U)\,$,
$t_{p'p}^{\downarrow\downarrow}(\mu_U)\,$. The matrix
$\hat{r}_{p'p}(\mu_U)$ is composed in a similar way. Taking into
account
that $\hat{\sigma}_{\nu}=\hat{\vec{\sigma}}\cdot \vec{\nu}$
$=\hat{\sigma}_i \nu_i $ and denoting
\begin{subequations}
\label{sp_cond_def}
\begin{gather}
g^L_{ji}=\frac{1}{2\pi \hbar}\sum_{pp'} \frac{1}{2}{\rm Tr}\left\{
\hat{t}_{p'p}^+(\mu_U)\,\hat{\sigma}_{j}\, 
\hat{t}_{p'p}(\mu_U)\,\hat{\sigma}_{i}
\right\}\\
\begin{split}
g^R_{ji}=\frac{1}{2\pi \hbar}\left[ \vphantom{\frac{1}{2}}\right.
\sum_{pp'}
\frac{1}{2} {\rm Tr} \left\{\hat{r}_{p'p}^+(\mu_U)\,
\hat{\sigma}_{\hspace{1mm}j}\,
\hat{r}_{p'p}( \mu_U)\,\hat{\sigma}_{\hspace{1mm}i} \right\} &
\\
-N_m(\mu_U)\delta_{ij}& \left.\vphantom{\frac{1}{2}} \right]
\end{split}
\end{gather}
\end{subequations}
we rewrite (\ref{J_j_L_R_t+_sigma_j_t_sigma_nu}) in the form
\begin{equation}
\label{sp_cur_g_L_R_ji}
\langle J_j\left(X,t=0\right)\rangle_{L,R}=\sum_i
\;g^{L,R}_{ji}\;\nu^{L,R}_i\;\delta \mu_{L,R}\, .
\end{equation}
We shall call the $3\times 3$ matrices $g^L$ and $g^R$ as spin
conductances. They determine the response of the spin current to
the
polarization of the left and right reservoirs, respectively.
For example, if the left reservoir is polarized along the 
$s$-th axis,
$\bm{\nu}^L=\bm{e}_s$ ($\bm{e}_s$ is the $s$-th coordinate ort), 
and 
the right reservoir is unpolarized, then according to
(\ref{J_J_L_J_R}) and (\ref{sp_cur_g_L_R_ji})
$\langle J_j\left(X,t=0\right)\rangle =g^L_{js}\delta \mu_L$ and
$g^L_{js}$ proves be the proportionality coefficient between
$\delta
\mu_L$ and $j$-th component of the spin current.

It is convenient to express the quantities $\nu^L_i \delta
\mu_L$
and $\nu^R_i \delta \mu_R$ in Eq. (\ref{sp_cur_g_L_R_ji}) in
terms
of 2D spin polarization densities $\langle {\cal
P}_{i}(XY)\rangle_L$
and $\langle {\cal P}_{i}(XY)\rangle_R$ corresponding to chemical
potentials of the left and right reservoirs. At small $\delta
\mu_L$
and $\delta \mu_R \ll E_F$ we have
\begin{equation}
\label{P_L_i_XY_expr_aver} \langle {\cal P}_i (XY)\rangle_{L/R}
=N_F
\nu^{L/R}_i \delta \mu_{L/R} \,,
\end{equation}
where $N_F=2\pi m^*/h^2$ is the 2D electron state density.
Substituting this expression into Eq.~(\ref{sp_cur_g_L_R_ji})
we rewrite the latter in the form
\begin{equation}
\label{J_j__LR_g_ji_LR_P_i_LR}
\langle J_j\left(X,t=0\right)\rangle_{L,R}= \frac{2\pi \hbar^2}{m^*}
\sum_i g^{L,R}_{ji}\langle {\cal P}_i (XY)\rangle_{L,R}\, .
\end{equation}
Combining (\ref{J_J_L_J_R}), (\ref{sp_cur_g_L_R_ji}), and
(\ref{J_j__LR_g_ji_LR_P_i_LR}) we obtain finally
\begin{multline}
\label{J_final}
\langle J_j\left(X,t=0\right)\rangle=
\langle J_j\left(X,t=0\right)\rangle_L+
\langle J_j\left(X,t=0\right)\rangle_R\\
=\sum_i \;g^L_{ji}\;\nu^L_i\;\delta \mu_L+\sum_i
\;g^R_{ji}\;\nu^R_i\;\delta \mu_R\\
= \frac{2\pi \hbar^2}{m^*}\left\{\sum_i g^L_{ji}\langle {\cal P}_i
(XY)\rangle_L +\sum_i g^R_{ji}\langle {\cal P}_i
(XY)\rangle_R\right\}\, ,
\end{multline}
This expression gives the spin current in the right lead in
terms of
the left and right reservoir polarizations.

Comparing our expression (\ref{sp_cond_def}) for the spin
conductance with the Landauer formula we see that they are
different
in the way that our spin conductance is written in terms of
both transmission and reflection coefficients. 
This difference is of
principal character since due to nonconservation of spin 
current we can not express the contribution
$\langle J_j\left(X,t=0\right)\rangle_R$ (containing the
reflection
coefficients) via the spin current in the left lead transmitted
from
the right.

\section{\label{transformation}Evolution of polarization along
$s$-th trajectory}

If we define the polarization of an electron as
\begin{equation}
\label{eq_pol}
{\bm P}={\rm Tr}\left\{\hat{\rho}\,\hat{\bm{\sigma}}\right\}\, ,
\end{equation}
then, according to Ref.~\onlinecite{Gottfr} the density matrix
describing its spin state may be represented in the form

\begin{equation}
\label{eq_dens_matr}
\hat{\rho}=\frac{1}{2}\left\{ \hat{\sigma_0}+{\bm P}\cdot
\hat{\bm{\sigma}}\right\} \, ,
\end{equation}
where $\hat{\sigma_0}$ is the  $2\times 2$ unit matrix. After
passing $s$-th trajectory, the spin density matrix is
transformed
into
\begin{equation}
\label{eq_dens_matr_transf}
\hat\rho'=\hat{S}_s\hat{\rho} \hat{S}_s^+ \, ,
\end{equation}
where $\hat{S}_s$ is the spin evolution operator given by Eq.
(\ref{expr_S}). Substituting (\ref{eq_dens_matr}) into
(\ref{eq_dens_matr_transf}) and then
(\ref{eq_dens_matr_transf}) into (\ref{eq_pol}) we obtain the
polarization of the electron
at
the end of the $s$-th trajectory
\[
{\bm P'}=\frac{1}{2}{\rm Tr}\left\{\hat{S}_s\left({\bm P}\cdot
\hat{\bm \sigma}\right)\hat{S}_s^+\hat{\bm \sigma}\right\}\, .
\]
Writing down this vector equation in components we obtain
\[
P'_i=K_{ij}^s P_j\, ,
\]
where $K_{ij}^s$ is given by Eq. (\ref{Kij}). Therefore, we see
that
$K_{ij}^s$ gives $i$-th component of the electron polarization
at
the end of $s$-th trajectory, provided that at its beginning the
electron had the unit polarization along the $j$-th axis.

\section{\label{char_times}Characteristic times}

In this Appendix we derive expressions for the particle
lifetime $\tau$,
the relaxation time associated with the finite width of the
ring $\tau_{\perp}$ and the characteristic time of one turn
$T_1$.
The widths of the ring ($d$) and leads ($W$) will be assumed
to be much less than the radius $a$.

Let us start with $\tau$. This time is determined by the
shortest of
two times: the mean time of particle escape from the ring and
the
dephasing time associated with inelastic electron-electron and
electron-phonon collisions. We will assume that the temperature
is
low enough to neglect the latter effect and will focus on the
escape
time. Any electron trajectory inside the ring is a set of
straight
segments. The probability that a current trajectory segment is
the
last one before escaping from the ring is $\tau_h/\tau$, where
$\tau_h$ is a time interval between two consecutive collisions
with
ring boundaries. On the other hand, the same probability may be
written as $2W/4\pi a$. Thence, $\tau\sim\tau_h(2\pi a/W)$. In
its
turn, $\tau_h$ can be estimated as
\begin{equation}
\label{tau_h}
\tau_h\simeq\frac{d}{v_F}\left<\left|\frac{1}{\cos\theta}\right|\right>_s\,
,
\end{equation}
where $\theta$ is the angle between the particle velocity and
the
radius-vector. Eq. (\ref{tau_h}) is valid for $\theta$ not too
close
to $\pi/2$, namely $|\pi/2-\theta| \gtrsim \sqrt{2d/a}$. The
average
in this equation is calculated assuming the isotropic
distribution
of $\theta$. A logarithmic divergence near $\theta = \pi/2$ is
removed by the cutoff $\sqrt{2d/a}$. We thus obtain
$\langle \left|1/\cos\theta\right|\rangle_s=(1/\pi) \ln(2a/d)$ 
and
\begin{equation}
\label{tau} \tau=2\frac{d}{v_F}\frac{a}{W}\ln(2a/d)\, .
\end{equation}

For evaluation of the winding time $T_1$ we introduce a
probability
$q$ that a particle changes its direction of motion along a 
ring arm 
after scattering from the ring boundary. 
In the case of diffusion
scattering $q=1/2$. If $q<1/2$, the specular reflection
prevails. In
such a situation the time
$\tau_h^{eff}\equiv\frac{1/2}{q}\tau_h$
plays a role of a mean free time for a particle that propagates
diffusively along a ring arm. The corresponding diffusion
coefficient $D$ can be evaluated as $\langle \Delta
x_{eff}^2\rangle_s/\tau_h^{eff}$, where $\langle\Delta
x_{eff}^2\rangle_s=(v_F\tau_h^{eff})^2\langle \sin^2\theta\rangle_s$
is the mean quadratic distance along a ring arm that an
electron passes during the time $\tau_h^{eff}$. Calculating the
average
$\langle \sin^2\theta\rangle_s$ in the same way as above we
obtain
\begin{equation}
\label{Delta_x_eff} D=\frac{dv_F}{4\pi q}\ln(2a/d)\,
.\end{equation}
The distance $L$ passed by a diffusing particle during the time
$T$
is $L=\sqrt{D T}$ and, for the winding number we
obtain,
accordingly
\begin{equation}
\label{w_2_av} \langle w^2\rangle_s=L^2/(2\pi a)^2=D
T/(2\pi
a)^2=T/T_1.
\end{equation}
Finally, we get from this equation and Eq. (\ref{Delta_x_eff})
\begin{equation}
\label{T_1} T_1\simeq 16\pi^3
q\left(\frac{a}{d}\right)^2\frac{d}{v_F \ln(2a/d)}\, .
\end{equation}

To find $\tau_{\perp}$ we use definition
(\ref{1_over_tau_perp}) together with Eqs.~(\ref{B_uu}) and
(\ref{B_psi_psi}). A simple
algebra gives
\begin{equation}
\label{tau_perp_expr} \tau_{\perp}= T_1
\frac{3}{4\pi^2}\frac{\zeta^2}{\gamma^4(1+\zeta^2)}
\left(\frac{a}{d}\right)^2\, .
\end{equation}

\end{document}